\documentclass[useAMS]{mn2e}
\usepackage{epsfig,natbib_vw,times}

\usepackage{amsmath,amssymb}

\bibpunct[, ]{(}{)}{;}{a}{}{,}

\def \aj {AJ}
\def \mnras {MNRAS}
\def \apj {ApJ}
\def \apjs {ApJS}
\def \apjl {ApJL}
\def \aap {A\&A}

\def \araa {ARAA}
\def \pasp {PASP}

\def \be {\begin{equation}}
\def \ee {\end{equation}}

\def\gsim{\mathrel{\lower0.6ex\hbox{$\buildrel {\textstyle >}
 \over {\scriptstyle \sim}$}}}
\def\lsim{\mathrel{\lower0.6ex\hbox{$\buildrel {\textstyle <}
 \over {\scriptstyle \sim}$}}}

\def\m@th{\mathsurround=0pt }
\def\eqalign#1{\null\,\vcenter{\openup1\jot \m@th
 \ialign{\strut\hfil$\displaystyle{##}$&$\displaystyle{{}##}$\hfil
 \crcr#1\crcr}}\,}

\def \Dz   {\Delta z}
\def \zabs {z_{\rm abs}}
\def \zqso {z_{\rm qso}}

\def \kms {\,km\,s$^{-1}$}

\def \civ {C~{\sc iv}}

\def \oiii {[O~{\sc iii}]}
\def \oii {[O~{\sc ii}]}

\def \mgii {Mg~{\sc ii}}
\def \feii {Fe~{\sc ii}}

\title[Intrinsic QSO absorbers]{Narrow associated QSO absorbers:
  clustering, outflows and the line-of-sight proximity effect}
\author[V. Wild et~al.]{
\parbox[t]{\textwidth}{\raggedright 
Vivienne Wild$^1$\thanks{vwild@mpa-garching.mpg.de}, 
Guinevere Kauffmann$^1$, 
Simon White$^1$,
Donald York$^2$,
Matthew Lehnert$^3$, 
Timothy Heckman$^4$, 
Patrick B. Hall$^5$,
Pushpa Khare$^6$,
Britt Lundgren$^7$,
Donald P.~Schneider$^8$,
Daniel Vanden Berk$^8$
}\\
\vspace*{6pt}\\
$^1$Max-Planck Institut f\"{u}r Astrophysik, Karl-Schwarzschild Str. 1,
85741 Garching, Germany \\
$^2$Department of Astronomy and Astrophysics and The Enrico Fermi Institute, University of Chicago,
Chicago, IL 60637, USA \\
$^3$GEPI, Observatoire de Paris, CNRS, University Paris Diderot; 5
Place Jules Janssen, Meudon, France \\ 
$^4$Center for Astrophysical Sciences, Department of Physics and
Astronomy, Johns Hopkins University, Baltimore, MD21218, USA \\
$^5$Department of Physics and Astronomy, York University, 4700 Keele
St.,Toronto, Ontario M3J 1P3, Canada\\
$^6$Department of Physics, Utkal University, Bhubaneswar 751004, India
\\
$^7$Astronomy Department, University of Illinois at Urbana-Champaign,
1002 West Green Street, Urbana, IL 61801, USA \\
$^8$Department of Astronomy and Astrophyscics, 504 Davey Laboratory,
University Park, Pennsylvania 16802, USA \\
}

\begin{document}
\maketitle

\begin{abstract}

Using data from the Sloan Digital Sky Survey data release 3 (SDSS DR3)
we investigate how narrow ($<$700km/s) \civ\ and \mgii\ quasar
absorption line systems are distributed around quasars. The \civ\
absorbers lie in the redshift range $1.6 < z < 4$ and the \mgii\
absorbers in the range $0.4<z<2.2$.  By correlating absorbers with
quasars on different but neighbouring lines-of-sight, we measure the
clustering of absorbers around quasars on comoving scales between 4
and 30\,Mpc. The observed comoving correlation lengths are $r_o\sim
5h^{-1}$Mpc, similar to those observed for bright galaxies at these
redshifts. Comparing with correlations between absorbers and the
quasars in whose spectra they are identified then implies: (i) that
quasars destroy absorbers to comoving distances of $\sim 300$kpc
(\civ) and $\sim 800$kpc (\mgii) along their lines-of-sight; (ii) that
$\ga 40$\% of \civ\ absorbers within 3,000\,km/s of the QSO are not a
result of large-scale clustering but rather are directly associated
with the quasar itself; (iii) that this intrinsic absorber population
extends to outflow velocities of order 12,000\,km/s; (iv) that this
outflow component is present in both radio-loud and radio-quiet
quasars; and (v) that a small high-velocity outflow component is
observed in the \mgii\ population, but any further intrinsic absorber
component is undetectable in our clustering analysis.  We also find an
indication that absorption systems within 3,000\,km/s are more
abundant for radio-loud than for radio-quiet quasars. This suggests
either that radio-loud objects live in more massive halos, or that
their radio activity generates an additional low-velocity outflow, or
both.

\end{abstract}


\section{Introduction}

Intervening absorption lines in QSO spectra provide a wealth of
information on the gaseous Universe from high redshift to the present
day.  They also allow us to probe the metallicity and ionisation state
of the gas in environments ranging from voids to galaxy halos and
disks.  QSO absorption line systems (QSOALSs) are attributed to two
main sources: (i) material associated with the host galaxy of the QSO
: either in radiatively driven winds close to the accretion disk
around the black hole or in outflowing material in the host galaxy and
its surroundings ; (ii) intervening galaxies along the
line-of-sight.

QSOALSs are usually split into three classes: (i) systems within a few
thousand km/s of the QSO systemic redshift, termed ``associated''
systems; (ii) systems with velocity differences in the range of 10,000
to 60,000\,km/s; (iii) systems beyond this velocity that are generally
assumed to be spatially disconnected from the QSO. Additionally,
QSOALSs are split into two different classes depending on their line
width. Broad absorption lines (BALs) are unambiguously associated with
outflowing material from the central region of the QSO; narrow
absorption lines (NALs), with typical line widths of several hundred
km/s, can be associated either with the QSO host or with intervening
systems.

It is generally agreed that strong (rest equivalent width
[EW]$\gsim$0.3\AA) metal NALs located at large velocity separations
from the QSO probe diffuse gas in the halos (and possibly disks) of
ordinary galaxies \citep[e.g.][]{1984A&A...133..374B,
1997ApJ...480..568S, 2007arXiv0709.1470T}.  The currently favoured
model postulates the existence of small (of order 1-10\,kpc diameter)
clouds distributed throughout a galaxy's dark matter halo \citep[see
e.g.][]{2007arXiv0706.4336C}.  The best studied transition is that of
\mgii$\lambda\lambda2796.4,2803.5$.  The imaging studies of
\citet{1993eeg..conf..263S} suggested that \mgii\ extends out to a
radius of $\sim$40$h^{-1}$\,kpc in galaxy halos. Recent results show
that \mgii\ absorbing gas extends to distances greater than 100\,kpc,
but with lower covering fraction \citep{2006ApJ...645L.105B,
2007ApJ...658..161Z, 2007arXiv0710.5765K}.

The picture becomes much more complicated as we look closer towards
the QSO in velocity space.  A substantial body of work has focused on
BALs, which cannot be associated with intervening galaxies, because
they have line widths of at least 1000\,km/s. Commonly reaching
sub-relativistic velocities of up to 30,000\,km/s BALs are
unambiguously associated with AGN-driven outflowing gas
\citep[e.g.][]{1979ApJ...234...33W,2003ApJ...593L..11Y,
2006astro.ph..3827R, 2007ApJ...665..990G,2007ApJ...656...73L}.

Many NALs are known to be caused by ordinary galaxies, and therefore
any additional population associated with the QSO must be studied
statistically. In general, one looks for an excess of NALs relative to
the background density in the vicinity of the QSO.  Because galaxies
are clustered, however, one expects an excess of NALs due to {\it
galaxies} surrounding the QSO. This excess is expected to extend out
to distances of a few galaxy correlation lengths ($\sim 20$ Mpc, or
velocity separations of $\sim$3000\,km/s). QSO absorbers at velocity
separations less than this nominal value of $\sim 3000$ km/s are
termed associated absorption lines (AALs).

Some authors have dealt with the clustered component by restricting
their analysis to velocity differences that are larger than this
value. For example, \citet{1999ApJ...513..576R} and
\citet{2001ApJS..133...53R} report an excess of narrow \civ\ absorbers
with velocities $>$5000\,km/s in radio quiet QSOs as compared to
flat-spectrum radio-loud QSOs. Other studies have attempted to gain
insight into the {\em physical nature} of AALs in order to ascertain
whether or not a proportion of these systems may be associated with
the QSO.  Based on line diagnostics and time variability arguments,
\citet{1997AJ....113..136B} and \citet{1997ApJ...488..155H,
2001ApJ...550..142H} conclude that some AALs lie in the QSO host
galaxy itself. Lower ionisation AALs probably lie at greater distances
from the QSO \citep{1997ASPC..128...13B}. \citet{vdb_assabs} find that
associated \mgii\ absorbers have higher ionisation than intervening
systems, clearly showing that ionising radiation from the QSO can
affect the state of the surrounding gas.  The fraction of AALs in
steep- and flat-spectrum radio-loud, and radio-quiet QSOs, may also
provide clues to the origin of the AALs \citep{1988qal..conf...53F,
1994ApJS...93....1A, 2001ApJ...549..133G,2002ApJ...568..592B,
2003ApJ...599..116V,2001ApJS..133...53R}.  However, the number of
radio-loud QSOs studied thus far has been quite small and the results
obtained have been inconclusive.

 There have been many studies of the clustering of NALs around small
samples of QSOs. These studies have yielded conflicting results and
conclusions
\citep{1979ApJ...234...33W,1986ApJ...307..504F,1982ApJS...48..455Y,1988ApJS...68..539S,
1994AJ....107.1219E}.  Recent studies with larger QSO catalogues have
generally found that there is an excess NAL population closely
associated with the QSO
\citep{2001ApJS..133...53R,2003ApJ...599..116V}, but there is still
ambiguity as to whether this population arises from neighbouring
galaxies or from gas associated with the QSO, its host galaxy and its
halo.

The nature of both AALs and higher velocity associated absorbers has
potentially important implications for understanding feedback
processes in QSOs. Feedback is a key ingredient in modern day galaxy
evolution models, invoked to explain observations as diverse as the
shape of the galaxy luminosity function, the apparent bimodality in
the colour distributions of galaxies, the entropy of the intracluster
gas, and the tight correlation between black hole and galaxy bulge
mass \citep[e.g.][]{2006MNRAS.365...11C, 2006MNRAS.370..645B}. The
basic principle is that a large source of energy, originating from
either an intense burst of star formation or an Active Galactic
Nucleus (AGN), causes the expulsion of gas from the galaxy. This
feedback halts star formation and heats the surrounding intergalactic
medium (IGM).

A few studies of both low- and high-redshift radio galaxies have
yielded direct observational evidence for significant (in terms of
total energy and mass) AGN-driven outflows
\citep{2005MNRAS.362..931E,2005A&A...444L...9M, 2007arXiv0710.1189M,
2006ApJ...650..693N,2007A&A...475..145N}, but such systems are very
rare.  BALs are observed in 15-20\% of all QSOs
\citep{1990BAAS...22..806F, 2003AJ....125.1784H, 2003AJ....126.2594R,
2006ApJS..165....1T}, but these are best understood in terms of winds
produced in the immediate vicinity of the black hole
\citep{1995ApJ...454L.105M,2000ApJ...545...63E}. The impact of such
winds on the intergalactic medium of the galaxy is not well understood
\citep{2002ASPC..255..329W}.

Associated narrow absorption lines are tantalizing, because they may
be the signature of galaxy-wide feedback
\citep{2007ApJ...663L..77T}. However, in order to reveal the true
outflow signal, the contribution from galaxies clustered around the
QSO must first be subtracted.  This can be done if one is able to
compute the cross-correlation between QSOs and absorbers on
neighbouring lines-of-sight, since such correlations cannot be due to
processes occuring within the QSO or its host galaxy.
\citet{2001MNRAS.328..805O} cross-correlated QSOs and metal absorption
systems detected in the 2QZ QSO survey. They detected a marginally
significant positive correlation on scales of 10$h^{-1}$\,Mpc.
\citet{2007ApJ...655..735H} used 17 Lyman Limit absorption systems
with transverse separations from the QSO of less than 6\,Mpc to
measure the clustering of optically thick neutral hydrogen around
QSOs. Their inferred clustering amplitude implied a 15-50\%
underdensity of line-of-sight absorption systems, likely caused by
photoevaporation of absorbers by the ionising flux of the QSO.

In this paper, we study the clustering properties of a sample of 6456
\civ\ systems and 16137 \mgii\ systems drawn from the SDSS Data Release 3
(DR3) QSO catalogue.  Our aim is to recover the fraction of AALs that
are truly intrinsic to the QSOs.  We perform a 3 dimensional
QSO-absorber cross-correlation analysis to measure the contribution to
the observed NALs from galaxies clustered around the QSOs.  We compare
and contrast our results for \mgii\ and \civ\ absorber samples. These
ions have very different ionisation energies, 15.03eV and 64.5eV
respectively, and thus provide information on the physical state of
the gas around QSOs.  In Section \ref{sec:sample} we describe the
absorber selection and the basic properties of our samples. Section
\ref{sec:los} presents the line-of-sight correlation between absorbers
and QSOs in redshift space. In Section \ref{sec:trans} we estimate the
large scale \mgii-QSO and \civ-QSO correlations.  In Section \ref{sec:model}
these 3-D correlations are used to estimate the excess number of
absorbers along the QSO line-of-sight due to clustering. The main
results are presented in Section \ref{sec:results}, including the
difference in NAL distributions for radio-loud and radio-quiet
QSOs. We conclude and summarise in Sections \ref{sec:disc} and
\ref{sec:summary}.

Throughout the paper we assume a flat cosmology with
$\Omega_\Lambda=0.7$, $\Omega_M=0.3$, $H_0=100\,h\,{\rm
km\,s}^{-1}\,{\rm Mpc}^{-1}$. Unless otherwise stated distances are
given in comoving units assuming $h=0.7$.


\section{The Sloan Digital Sky Survey}

The SDSS \citep{2000AJ....120.1579Y} is using a CCD camera
\citep{1998AJ....116.3040G} on a 2.5-m telescope
\citep{2006AJ....131.2332G} at Apache Point observatory to perform a
5-band \citep{1996AJ....111.1748F} photometric and spectroscopic
survey of the high Galactic latitude sky.  Photometric calibration is
provided by simultaneous observations with a 20-inch telescope at the
same site \citep[see][]{2001AJ....122.2129H, 2002AJ....123.2121S,
2002AJ....123..485S, 2006AN....327..821T}. The survey data-processing
software measures the properties of each detected object in the
imaging data, and determines and applies photometric calibrations
\citep{2003AJ....125.1559P, 2004AN....325..583I}. The spectroscopic
survey provides medium resolution (R$\sim$2000) spectra with a
wavelength coverage from 3800 to 9200\AA; details can be found in
\citet{2000AJ....120.1579Y} and \citet{2002AJ....123..485S}.
Spectroscopic targets chosen by the various SDSS selection algorithms
are arranged onto a series of 3'' diameter circular fields
\citep{2003AJ....125.2276B}. QSO targets are selected on colour and
$i-$band point spread function (PSF) magnitude; for details see
\citet{2002AJ....123.2945R} and \citet{2007AJ....134..102S}. QSO
redshifts are predominantly derived from a fit to the QSO template of
\citet{2001AJ....122..549V}. 

The third data release (DR3) of the spectroscopic survey covers 3732
sq. degrees \citep{astro-ph/0410239}. The DR3 QSO catalogue
\citep{2005AJ....130..367S} provides a more complete sample of QSOs
than the SDSS standard pipeline and corrects the redshifts of a small
number of objects.


\section{The absorption line samples}\label{sec:sample}

The absorption line samples used in this paper are based on the
absorption line catalogue described by York et~al. (2006) and York et
al. (2008, in preparation).  The catalogue contains a list of narrow
($\lsim$700\kms) absorption features detected in QSO spectra drawn
from the SDSS DR3 QSO catalogue \citep{2005AJ....130..367S}.
Independent searches are carried out for both the
\civ$\lambda\lambda1548.2,1550.8$ and
\mgii$\lambda\lambda2796.4,2803.5$ doublets, and the equivalent width
(in the QSO frame) and significance of individual line detections are
catalogued. We select \civ\ absorbers with equivalent width of the
strongest line in the doublet, $\lambda1548.2$, greater than 0.3\AA.
We apply the same equivalent width cut to the \mgii\ doublet.

In this paper, we retain only those systems with detection
significance $>4\sigma$ for the primary line, and $>3\sigma$ for the
secondary line.  In each sample, duplicate entries are also
removed. These arise because the master catalogue contains the result
of independent searches for \mgii, \civ, and \feii.  We remove
duplicates with exactly the same $\zabs$, $\zqso$, and line equivalent
widths (EWs). If there are two absorbers with redshifts within 0.005
of each other (corresponding to $\Delta\lambda_r$ of 7.7\AA\ for \civ\
and 14\AA\ for \mgii), the strongest system is retained.

The absorber sample is then cross-matched with the SDSS DR3 {\it
specobj} view in order to obtain a well-defined QSO sample with
photometric and spectroscopic measurements. The FIRST radio survey
fluxes are extracted from the DR3 QSO catalogue
\citep{1995ApJ...450..559B}.  This gives a total of 15416 and 24406
\civ\ and \mgii\ absorption systems.

Finally, the SNR of the QSO spectra must be sufficient to allow lines
above a certain equivalent width to be detected throughout their
wavelength coverage, otherwise we may introduce artifacts into our
correlation functions.  Monte Carlo techniques could in principle be
employed to simulate the full selection function of an absorption line
search. However, with the large number of absorbers and QSOs available
to us from the SDSS we can afford to make stringent cuts to ensure a
(roughly) uniformly selected sample.

\begin{figure}
\includegraphics[scale=0.5]{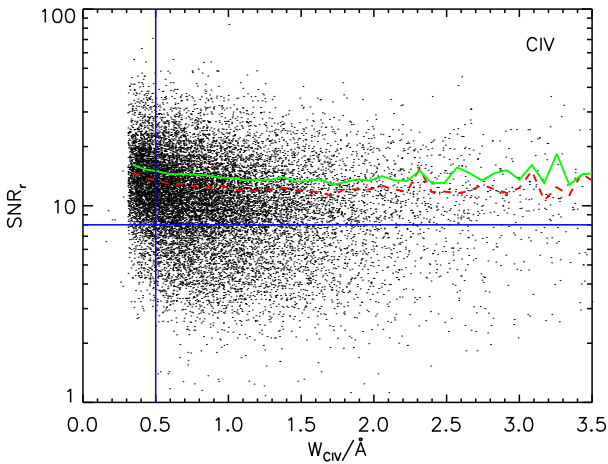}
\caption{$r$-band SNR of the background QSO spectrum vs. EW of
  detected \civ$\lambda1548$ lines. Straight blue lines show the
  sample cuts used when measuring the cross-correlation of absorbers
  with QSOs. The dashed red and solid green lines show the mean and
  median SNR as a function of EW for QSOs above the horizontal blue
  line. }\label{fig:sample}
\end{figure}

We therefore remove absorbers which fall below 4000\AA\ or above
9000\AA\ observed frame wavelength to avoid the noisiest regions of
the SDSS spectra. Figure \ref{fig:sample} shows the $r$-band SNR of
the QSO spectra vs.  EW of the \civ\ absorption lines (computed in the
absorber rest-frame).  There is a sharp rise in the median and mean
SNR of the spectrum as the EW of the absorber drops below 1\AA:
i.e. weak absorbers are preferentially found in high SNR spectra.
Cutting at per-pixel-SNR$>8$ (where a pixel covers 69\,km/s in
velocity space) and EW$>0.5$\AA\ gives roughly constant mean and
median SNR as a function of absorber EW (red and green lines). We
therefore adopt these cuts, reducing our sample to 7393 \civ\ and
16173 \mgii absorbers.

We have chosen not to eliminate BALQSOs from our sample, as has been
done in most previous studies. Our reason for retaining them is that a
BALQSO can also show narrow absorption at a different velocity to the
broad absorption. As the physical process responsible for the
presence of a BAL is not well understood, we do not wish to bias the
NAL sample by removing QSOs with BALs from our sample
entirely. Unfortunately the absorption line catalogue suffers from a
large number of ``false positive'' line identifications in the regions
of the BAL troughs.  BAL troughs are more often observed in \civ\ than
in \mgii\, so we visually inspected all \civ\ absorber candidates to
eliminate false positive detections, leaving us with 6456 \civ\
absorbers. There is a continuum of line widths from NALs through to
BALs; the criteria imposed during this visual inspection were that the
lines had velocity widths smaller than 700\,km/s and both members of
the doublet were clearly identifiable as an individual system in
velocity space, rather than being part of broader velocity
structures. Except for the most saturated cases, the two lines of the
doublet are clearly distinguishable using this velocity width
limit. As we do not determine the completeness of our absorber sample
for every QSO spectrum, the inclusion of the BALs will cause us to
underestimate number densities slightly where the flux in
the BAL troughs reaches zero. However, the fraction of BALQSOs is
around 10-15\% \citep{2003AJ....125.1784H,2006ApJS..165....1T} and
their typical line widths are one to a few thousand km/s. Thus, the
fraction of the total pathlength lost due to BAL troughs is small.

For the larger \mgii\ sample, only absorbers within
$v/c<0.015$ of the background QSO were inspected, and only  false
positive identifications caused by BAL troughs and poor continuum fits
around the QSO emission lines were removed. This procedure leaves
16137 \mgii\ absorbers. In fact, the majority of false positive line
identifications in the \mgii\ sample are caused by OH skylines
\citep[see][]{2005MNRAS.358.1083W}. 

\begin{figure*}
\includegraphics[scale=0.5]{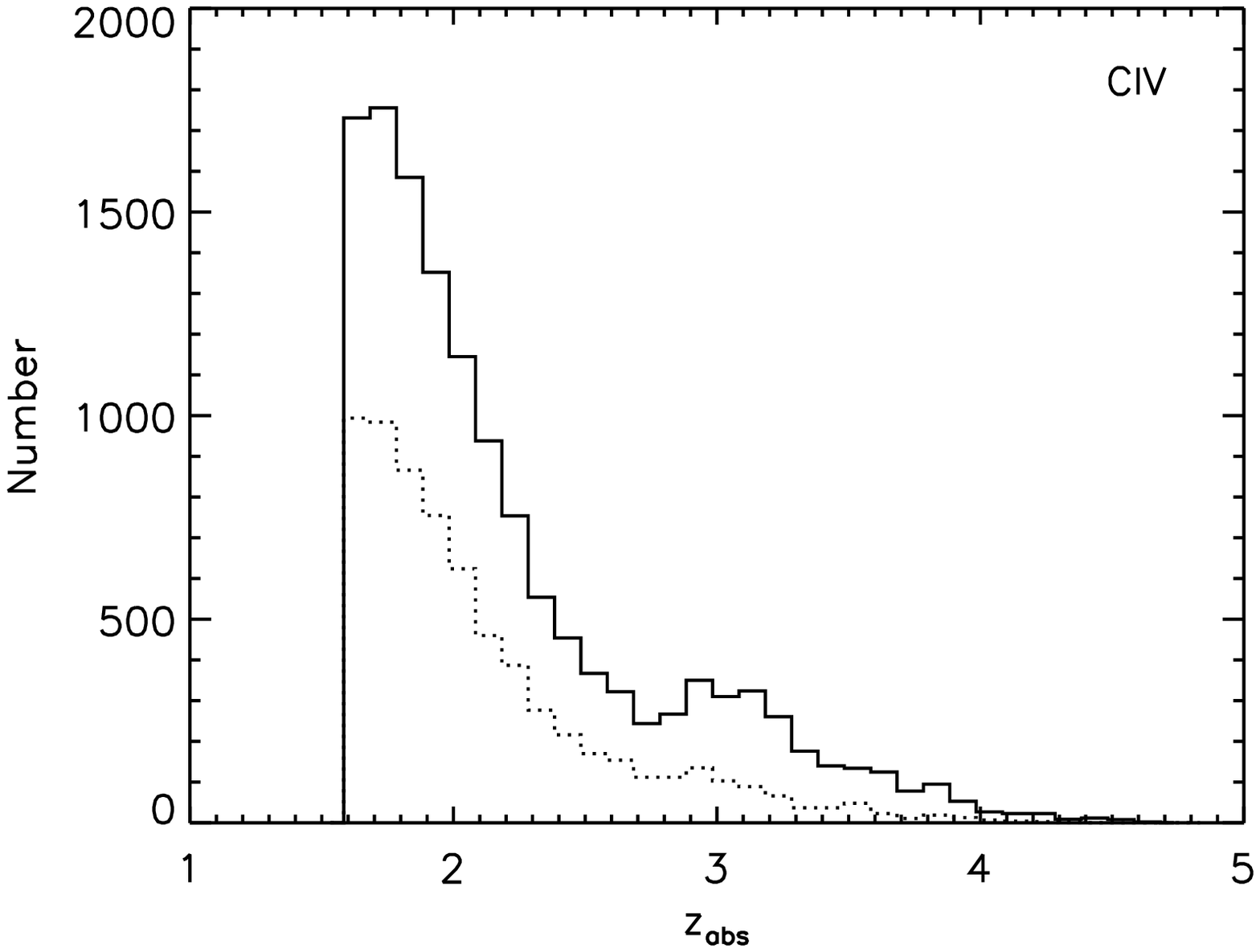}
\includegraphics[scale=0.5]{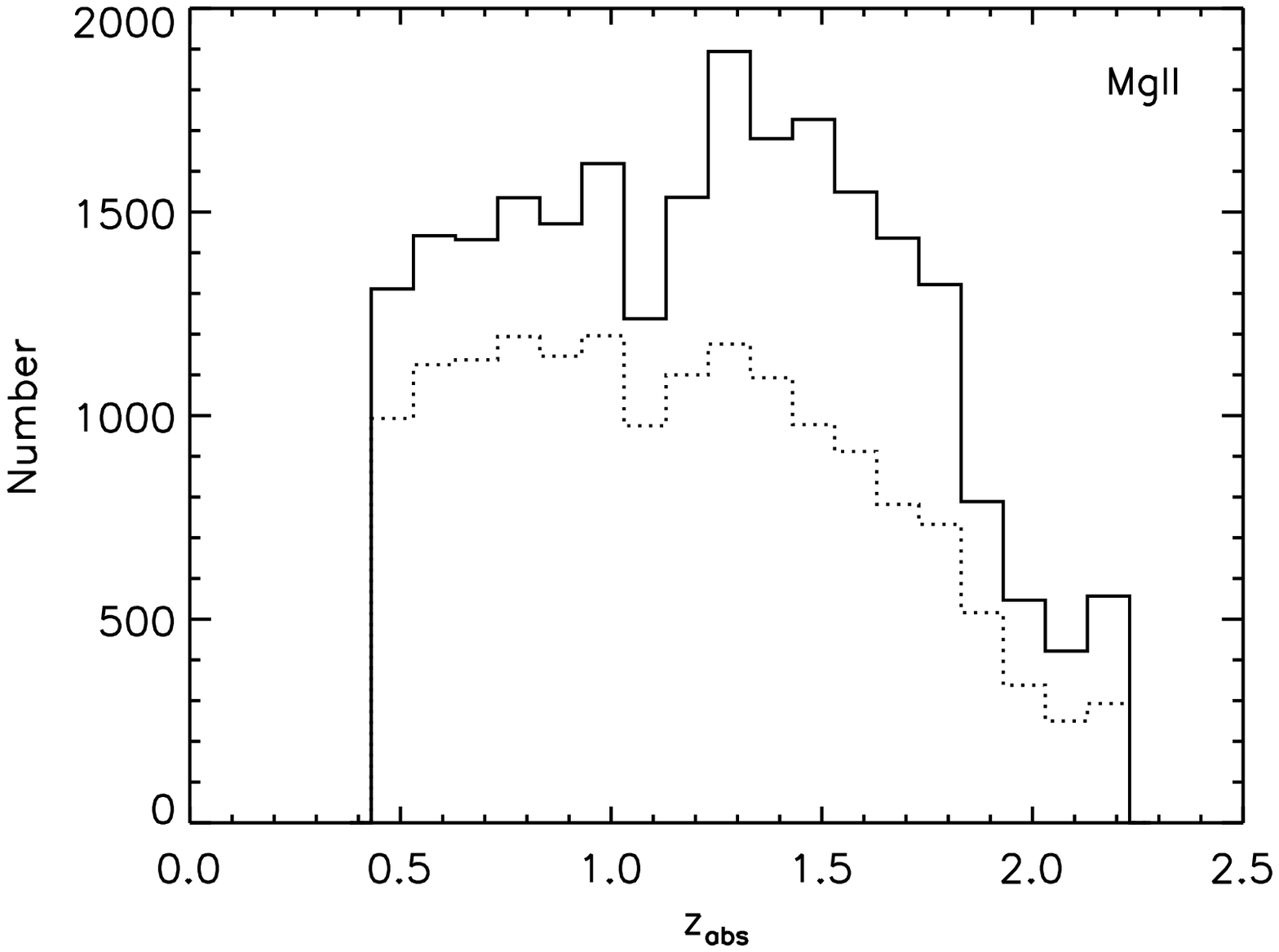}
\caption{Left: The redshift distribution of the \civ\ absorbers before
  (full) and after (dotted) the cuts on QSO spectrum SNR and
  absorber EW. Right: The same for the \mgii\ sample. }\label{fig:raw}
\end{figure*}

Our samples of \civ\ and \mgii\ absorbers have mean redshifts of 2.08
and 1.17 respectively.  Figure \ref{fig:raw} shows the observed
redshift distribution of the \civ\ and \mgii\ samples, before and
after the SNR and EW cuts are made. The cuts do not appear to
significantly bias the redshift distributions.

\section{Line-of-sight distribution of QSO-absorber separations}\label{sec:los}

\begin{figure*}
\includegraphics[scale=0.5]{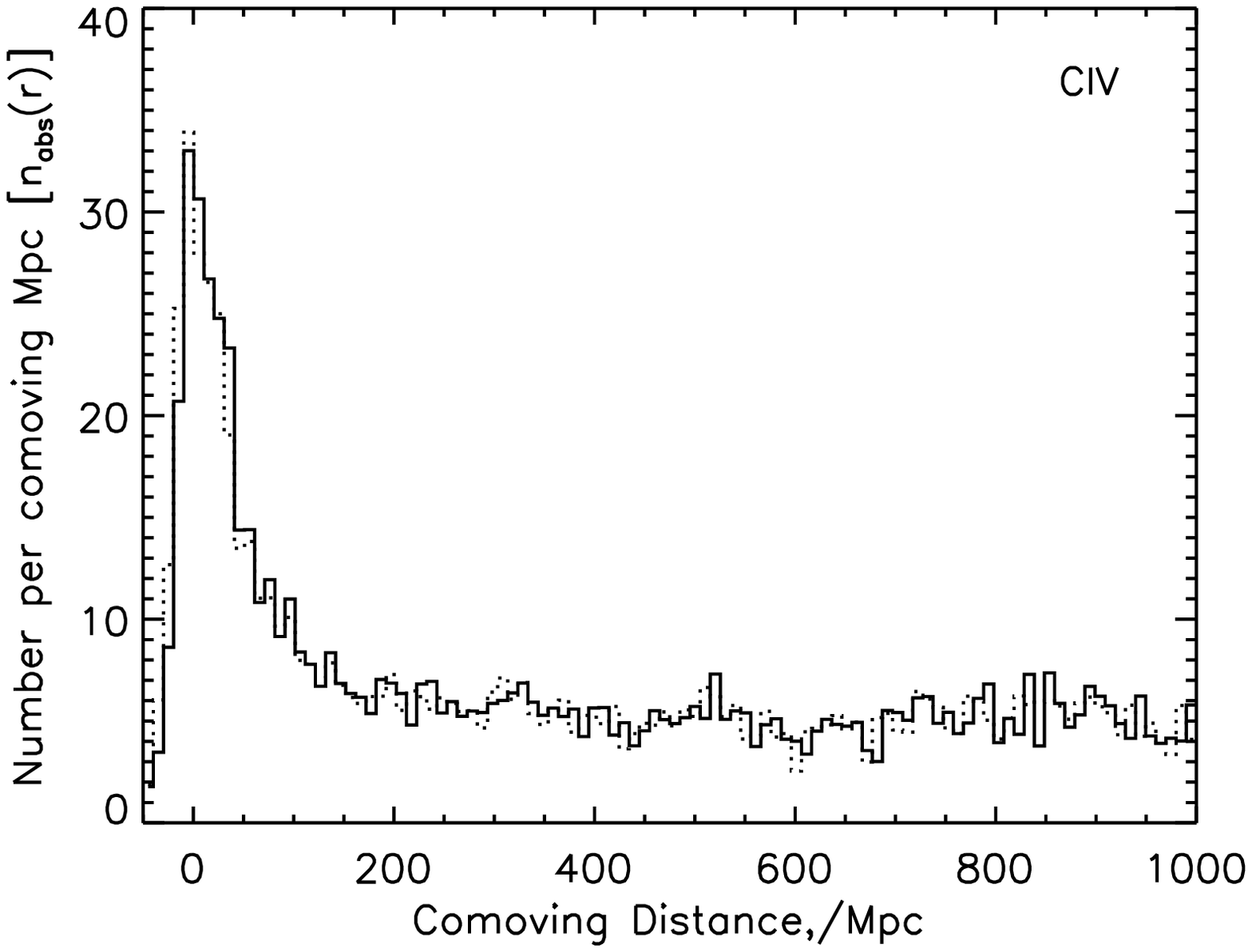}
\includegraphics[scale=0.5]{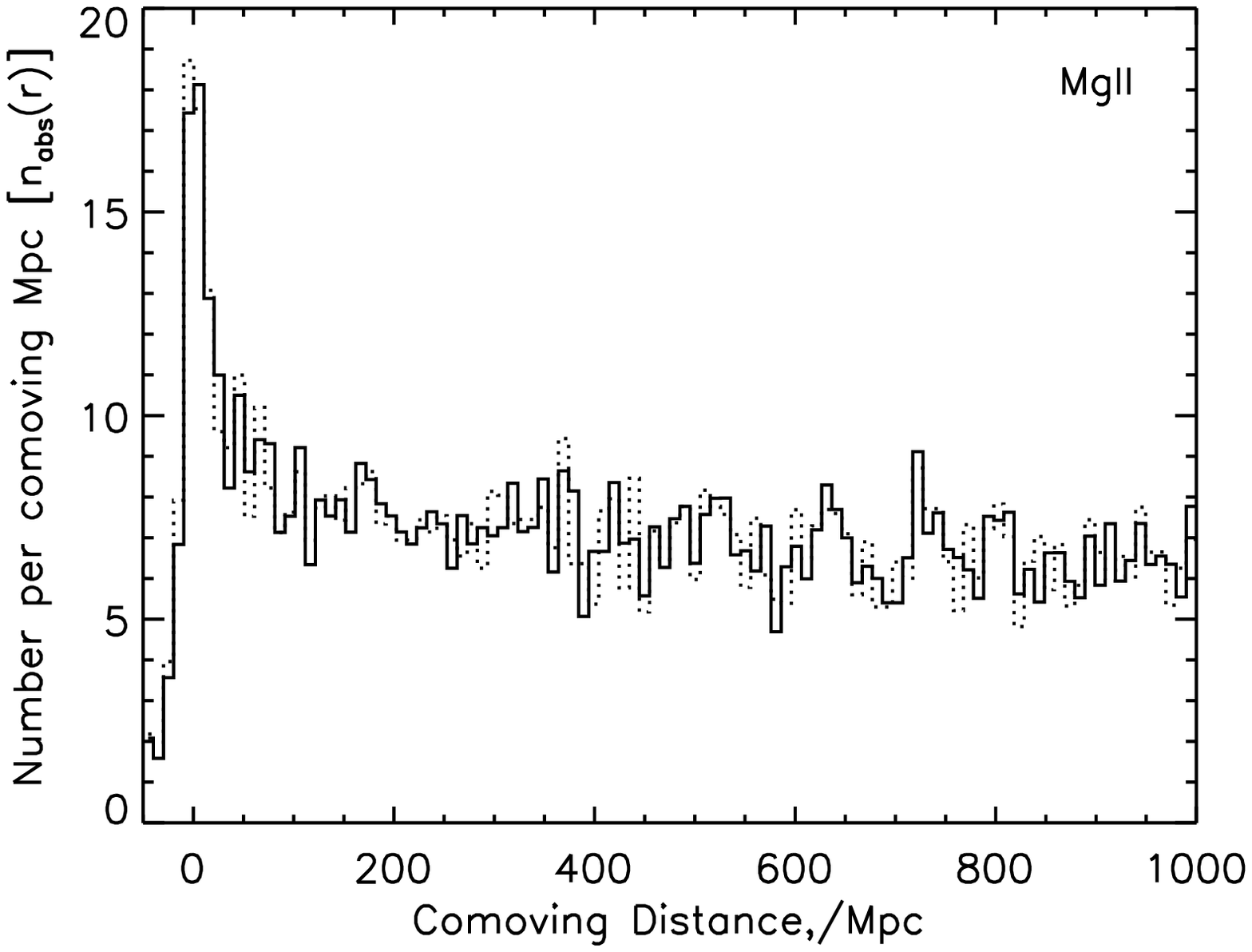}
\caption{{\it Left}: The number density of line-of-sight \civ\ absorbers found
  in QSO spectra as a function of comoving distance from the QSO,
  after (full) and before (dotted) correction of the QSO redshifts for
  the bias caused by blue-shifted UV emission lines.  The distribution
  has been corrected for completeness due to the decreasing
  pathlengths available to find an absorber as distance
  increases. Note the peaked excess of absorbers lying very close to
  the QSO, the significant number of negative velocity absorbers, and
  tail of excess absorbers extending to inferred distances of around
  150\,Mpc. {\it Right}: The same for an independent sample of \mgii\
  absorbers. An excess is again observed close to the QSO; a
  small tail to larger distances may also be present. }\label{fig:dz}
\end{figure*}

To get a first impression of the quality of our data, Figure
\ref{fig:dz} shows the number density of \civ\ and \mgii\ absorption
line systems as a function of comoving distance ($r$) from the
background QSO. In both cases, the distributions have been corrected
for the fact that the pathlength available to find an absorber
decreases as comoving distance increases.  The general form of these
distributions is robust to even significant changes in the sample.
Here, we choose to present line-of-sight distributions in comoving
distance rather than the more commonly used velocity parameter
$\beta\equiv v/c$, as we aim to quantify the distribution of absorbers
caused by galaxy clustering.  We will return to the $\beta$
distributions in Section \ref{sec:results}.

We note that there are clear similarities between the line-of-sight
distributions of the \mgii\ and \civ\ absorber samples. After
correction for survey completeness, both show a near constant absorber
density at large distances from the QSO, as we would expect for an
intervening absorber population. They also both display a very
pronounced enhancement in the fraction of absorbers within 25-50\,Mpc
from the QSO.  The \civ\ absorbers in addition show an enhanced number
of absorbers out to 150\,Mpc from the QSO. A similar, although much
smaller enhancement may be present out to 75\,Mpc in the \mgii\
absorbers.

As will be described more fully in the following sections,
the observed distributions may be interpreted as the superposition of
(i) a narrow spike at $r<50$Mpc due, in
part at least, to galaxy clustering; (ii) a tail 
to high velocities caused by sub-relativistic,
moderately ionised outflows driven by the QSO (seen primarily
in the \civ\ distribution ); (iii) a constant
background level of intervening absorbers.

Finally, it has been claimed that systematic uncertainties in the
estimated redshifts of the higher redshift QSOs in our sample could
potentially lead to false conclusions being drawn about the nature of
the associated absorbers \citep{2002AJ....124....1R}.  The full effect
of such systematic line shifts on the SDSS redshifts derived by
fitting a QSO template to the SDSS spectra is complex. A detailed
analysis of the full redshift-dependent systematic shifts will be
presented in Hewett \& Wild (2008, in preparation). A preliminary
analysis comparing the line-derived redshifts to the quoted SDSS
redshifts leads us to shift QSOs with $1.1<z<2.1$ redwards by
200\,km/s and QSOs with $z>2.1$ redwards by 415\,km/s. Although shifts
of this order of magnitude appear warranted by the present dataset,
the precise values used make only small differences to our final
results, and are likely to be refined by future work. The dotted
histogram in Figure \ref{fig:dz} shows the line-of-sight NAL
distribution before correcting the QSO redshifts for this bias, while
the solid histogram shows the corrected distribution. The difference
is very small.

\section{3-D correlation of absorbers and QSOs}\label{sec:trans}

In this section we
cross-correlate the absorbers with QSOs {\it on neighbouring
lines-of-sight}, in order to estimate the contribution to the
line-of-sight absorber distribution from  galaxies clustered around
the QSO hosts.  

It is generally believed that QSOs reside  in galaxies
\citep[e.g.][]{1973ApJ...179L..61K,1993MNRAS.264..455D}.  Strong
metal absorbers also arise primarily in ordinary galaxy halos and disks
\citep{1969ApJ...156L..63B, 1978ApJ...220...42B, 1997ApJ...480..568S,
2007ApJ...658..161Z}. Because galaxies are clustered, we expect to see
some enhancement in the number of absorbers close to the QSO.   
With the high density  of QSOs  available in the SDSS survey,
one can estimate this enhancement  by counting the number of 
absorbers along lines-of-sight that pass close to each QSO.

The standard measures of galaxy clustering are the 
2-point auto- and cross-correlation functions. These
are typically well approximated by power laws:
\begin{equation}
\xi(r) = \left( \frac{r}{r_0} \right)^{-\gamma}\label{eq:xi}
\end{equation}
where $\gamma\approx1.8$ and $r_0\approx5h^{-1}$Mpc for bright
galaxies at low redshifts \citep{1980lssu.book.....P}.  The ratio of
the number density of pairs separated by distance $r$ to that expected
for a random and uniform distribution of objects with the same number density is
$1+\xi$.  In this section, we estimate $\xi_{QA}$, the QSO-absorber
cross-correlation function. \footnote{We note that in reality we
measure the redshift-space cross-correlation function. We do not
attempt to correct for redshift space distortions: our approach is
conservative because on small scales, $\lsim10$\,Mpc, non-linear
$z$-space distortions will act to reduce the observed clustering
amplitude in a similar way to the effect of errors on the QSO
redshifts. On larger scales, the effect of linear $z$-space
distortions on the correlation function is small at the redshifts of
interest here \citep[see e.g.][for a detailed
discussion]{2005MNRAS.356..415C}.}.

\subsection{3-D correlation estimation}

\begin{figure*}
\includegraphics[scale=0.3]{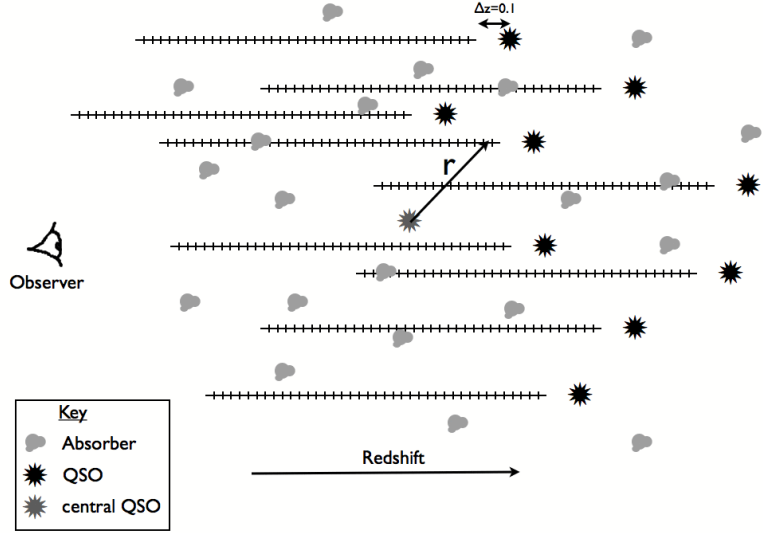}
\caption{A pictorial representation of the method used to calculate
the 3-D cross correlation between QSOs and absorbers. From a central
QSO we count, using small bins of $\Delta z$, the observed and
expected number of absorbers along neighbouring sightlines to
background QSOs. These numbers are accumulated as a function of the
comoving separation, $r$, between the center of the $\Delta z$ bin and
the central QSO. The process is repeated for all central QSOs, thereby
building a 3-D cross-correlation estimate. }\label{fig:method}
\end{figure*}

The basic principle of our estimation method is to compute the excess
number of QSO-absorber pairs relative to an unclustered sample, as a
function of comoving separation.  The number of QSO-absorber pairs
expected for a uniformly distributed sample is determined by computing
the number of absorbers expected at random along the actual sightlines
used to find the observed pairs. In this way, we automatically account
for selection biases such as those caused by fibre collisions in the
SDSS survey. Our final measurement yields the excess number of
absorbers as a function of comoving distance from a QSO. We can then
use this measurement to ascertain whether the excess seen along the
line-of-sight is consistent with galaxy clustering alone, or whether
some component of the excess is intrinsic to the QSO itself. The
scheme is presented pictorially in Figure \ref{fig:method}, to which
we will refer throughout this Section.

First we build our QSO-QSO pair catalogue. For each QSO in the DR3 QSO
catalogue, the SDSS Catalogue Archive Searcher (CAS) is used to find
all QSOs at higher redshift and within 3 degrees of it on the sky. We
cross-match the DR3 QSO catalogue with the {\it specobj} view of the
CAS in order to ensure a unique sample of QSOs from main SDSS survey
plates. The DR3 QSO catalogue contains 45,988 QSOs; this results in
$\sim$6.5 million pairs above a redshift of 0.3. Only 10823 DR3 QSOs
have SNR above the threshold (8 per pixel), and $\zqso>1.58$ so that
\civ\ can be seen above a restframe wavelength of 4000\AA. This
reduces the number of QSO-QSO pairs drastically for the \civ\ analysis
compared to the \mgii\ analysis. QSOs below the SNR threshold are
retained only as central QSOs (see Figure \ref{fig:method}) in the
correlation analysis, i.e., the sightlines provided by their spectra
are not included in the absorption system search or in the control
absorber counts.

Next, we select the sightlines which we will use in the analysis.  We
only consider sightlines from QSOs above the SNR threshold. As
indicated in Figure \ref{fig:method} we restrict the sightlines to
start at $\Dz>0.1$ from their QSO host, and to lie redwards of 1250\AA\
in the QSO rest frame (to avoid the strong Lyman-$\alpha$ line at
1215\AA\ and the Lyman-$\alpha$ forest bluewards of
Lyman-$\alpha$). The velocity difference criterion $\Dz>0.1$ ensures
that the final signal is not affected by the excess of associated
absorbers with sub-relativistic velocities described in Section
\ref{sec:los}.  Because of the restriction that sightlines must end at
1250\AA\ we cannot increase this limit much further without
drastically reducing the number of QSO- \civ\ absorber
pairs. Contained within these sightlines are 3374 \civ\ and 13504
\mgii\ absorbers.

\begin{figure*}
\includegraphics[scale=0.5]{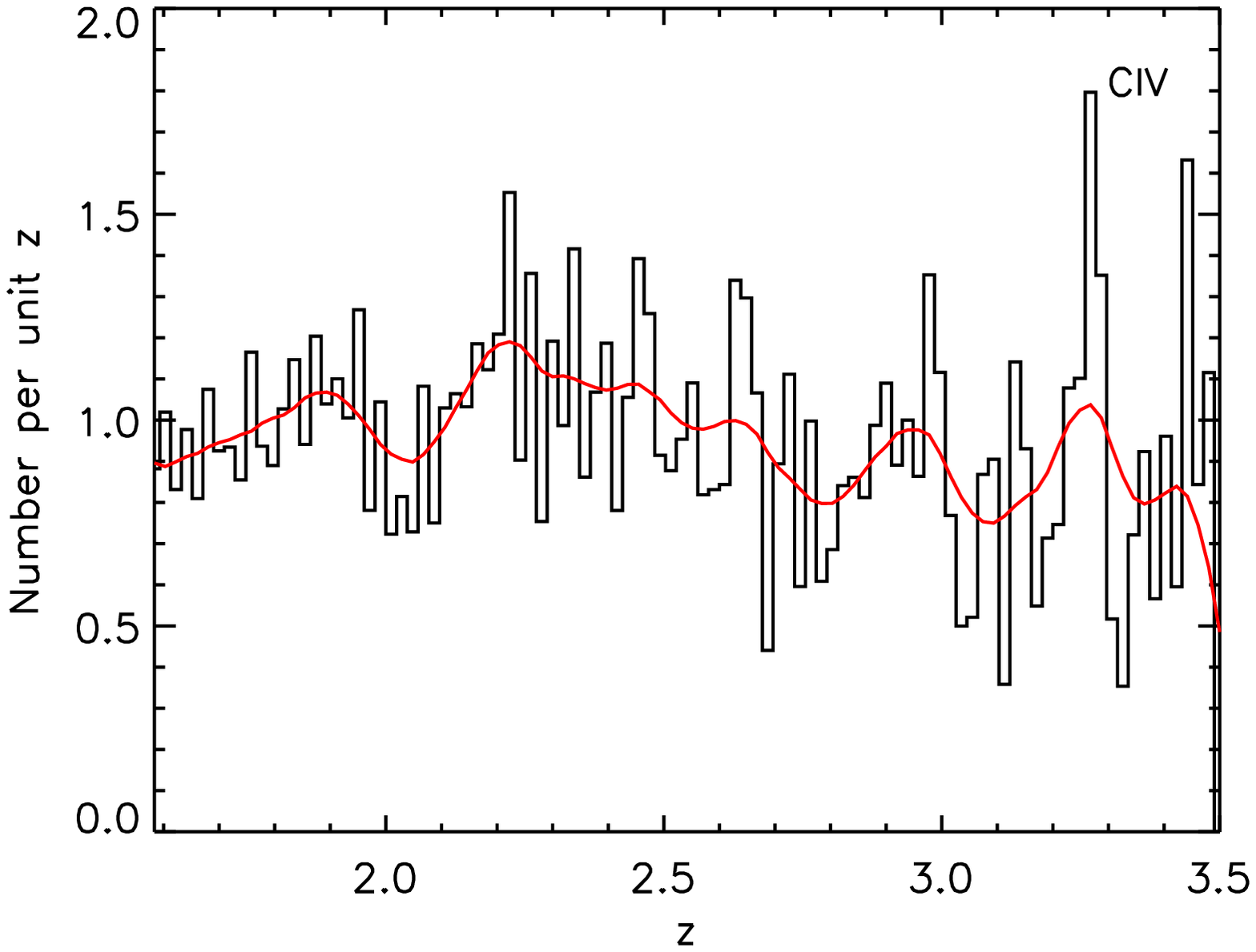}
\includegraphics[scale=0.5]{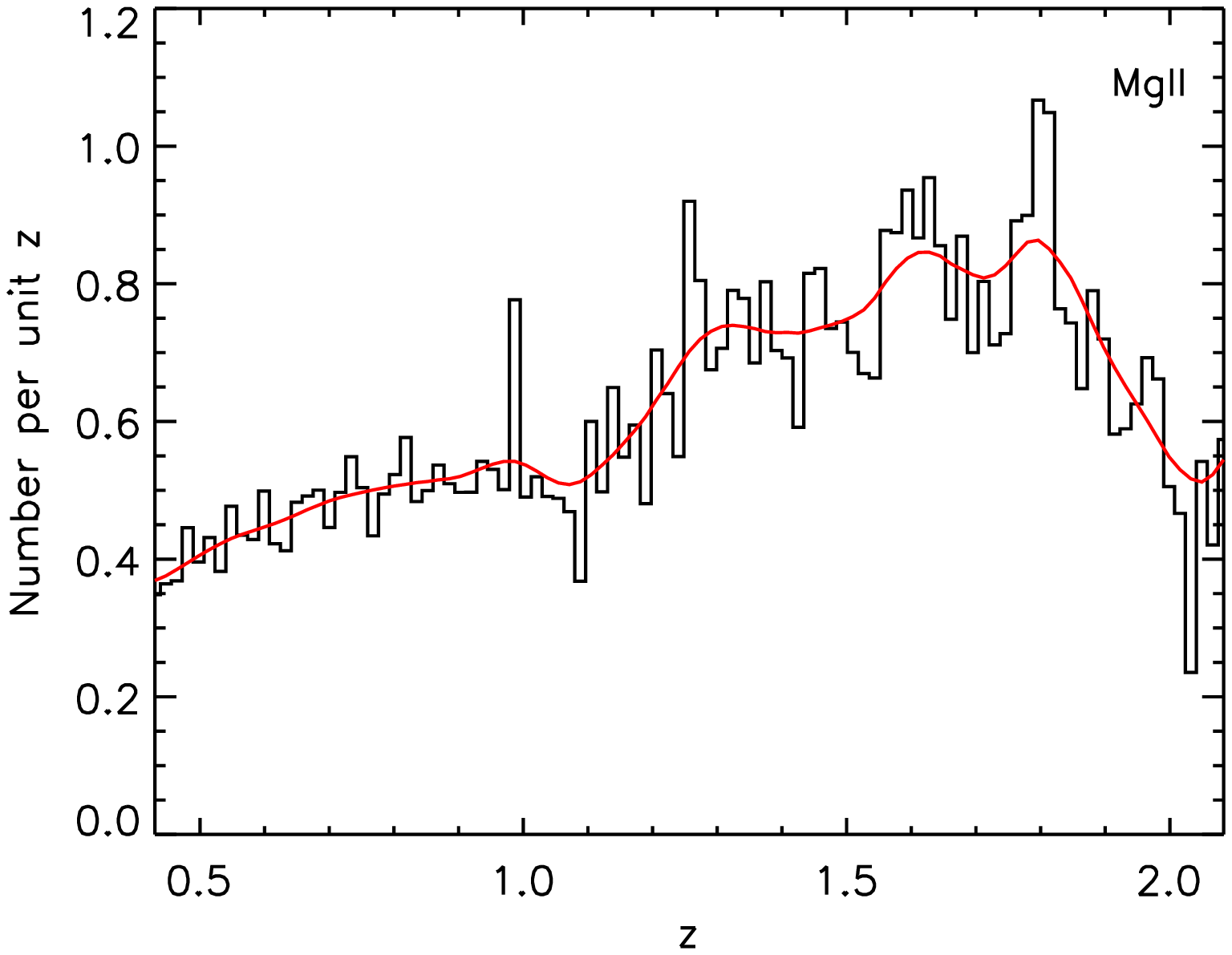}
\caption{{\it Left:} Number density of detected \civ\ absorbers as a
  function of $\zabs$, $n$($z$). Overplotted in red is the smoothed
  distribution used in the analysis. {\it Right:} Same for \mgii\
  absorbers. }\label{fig:nz}
\end{figure*}

Using these sightlines and absorbers, we calculate the number density
of absorbers as a function of redshift $n$($z$) shown in Figure
\ref{fig:nz}, and use this to estimate the expected number of
absorbers in each $\Delta z$ bin along each sightline. Then, for each
central QSO we use the sightlines of all its paired QSOs to measure
the expected and observed number of absorbers as a function of
comoving separation between the central QSO and the position along the
sightline\footnote{In a flat universe, the comoving distance between
two objects is given by the square root of the sum of the squares of
the angular and line-of-sight comoving distances.}. In Figure
\ref{fig:method} the arrow, labelled $r$, indicates the comoving
distance between the central QSO and one of the bins along one of the
sightlines. In this bin, no absorber is found so no addition is made
to the observed number of absorbers at this comoving
separation. However, a small addition will be made to the expected
number of absorbers at this separation, based on the pre-calculated
$n$($z$) at the redshift of this $\Delta z$ bin. As well as accruing
the observed and expected numbers of absorbers as a function of
comoving separation, we sum the number of contributing sight lines,
which is necessary in the estimation of the errors. Finally, because
we must bin our results in $\Delta r$, and we do not have enough
sightlines to create infinitesimal bins, we must account for the fact
that more sightlines within a $\Delta r$ range will lie towards the
outer edge of the range than the inner edge i.e. the effective radius
of our $\Delta r$ bin is not centered on to the bin, but biased
towards the outer edge. We do this by accumulating the expected number
of absorbers, weighted by $r^{-1.7}$ (see Equation \ref{eq:reff}).

The final 3-D correlation estimate is given by:
\begin{equation}\label{eq:corrnfn}
1+\xi({\rm r_{eff}}) = \frac{\rm N_o(r_{eff})}{\rm N_e(r_{eff})}
\end{equation}
where ${\rm N_o(r_{eff})}$ is the observed number of absorbers in a
$\Delta r$ bin with effective radius ${\rm r_{eff}}$, and ${\rm
N_e(r_{eff})}$ is the expected number as described above.  The
effective radius of each of the $\Delta r$ bins is estimated to be:
\begin{equation}\label{eq:reff}
{\rm r_{eff}} = \left( \frac{\sum {\rm N_e} {\rm r}^{-\gamma}}{\rm N_e} \right)^{(-1/\gamma)}
\end{equation}
where the sum is over all $\Delta z$ bins (segments of the sightlines
in Figure \ref{fig:method}) that contribute to the $\Delta r$
bin. $\gamma$ is chosen to be 1.7, which is close to the measured
exponent in the \mgii-QSO correlation (see below), although the
precise value used does not change the results significantly.

On small scales the QSO-absorber pairs in each $\Delta r$ bin are
independent (in general, absorbers are only paired to one QSO on small
scales), thus the error is estimated assuming Poisson statistics:
\begin{equation}
\Delta\xi({\rm r}) = \frac{1+\xi({\rm r})}{\sqrt{\rm N_o(r)}} =
  \frac{\sqrt{\rm N_o(r)}}{\rm N_e(r)}.
\end{equation}
On larger scales the pairs in each $\Delta r$ bin are no longer
independent (one absorber can be paired with many QSOs in the same
bin) and the Poisson errors underestimate the true errors. Following
\citet{1994MNRAS.271..753S}, as the number of pairs approaches the
number of absorbers in the analysis (N$_{\rm abs}$), the errors are
approximated by:
\begin{equation}
\Delta\xi({\rm r}) = \frac{1+\xi({\rm r})}{\sqrt({\rm N_{abs}})}.
\end{equation}

Finally, we calculate the significance of a positive clustering signal (excess
number of pairs) assuming a Poisson distribution. Assuming no
clustering, the probability of observing N$_{\rm o}$ or more pairs at
any given $r$ is given by:
\begin{equation}\label{eq:binomial}
P(k \ge {\rm N_o}) = \sum_{k=N_o}^\infty\frac{N_{\rm
    e}^k e^{-N_{\rm e}}}{k!}.
\end{equation} 
In the closest bins with fewest observed pairs, we checked that the
probabilties derived here agree with those derived from a Binomial
distribution.

\subsection{Clustering detection}

\begin{figure*}
\includegraphics[scale=0.5]{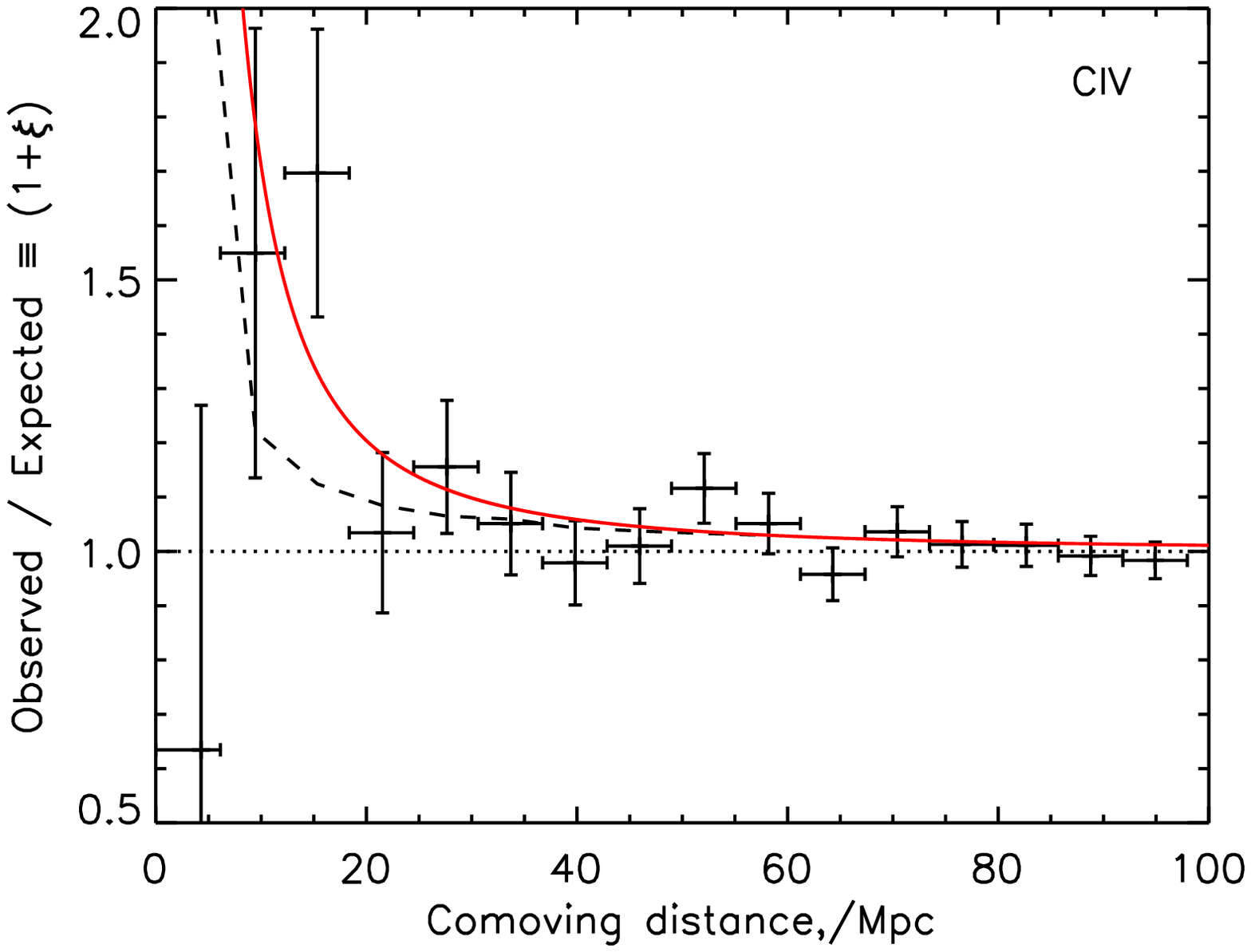}
\includegraphics[scale=0.5]{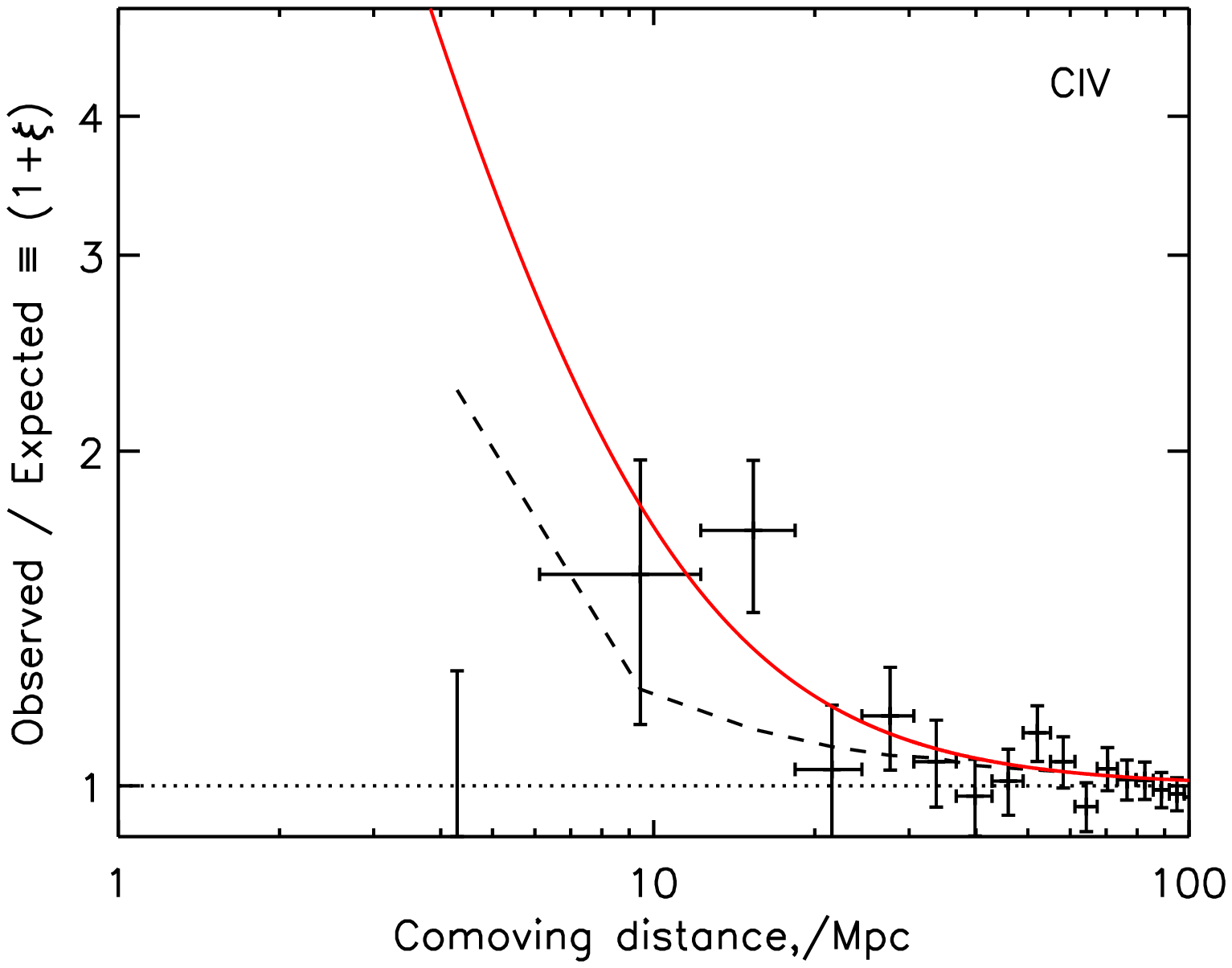}\\
\includegraphics[scale=0.5]{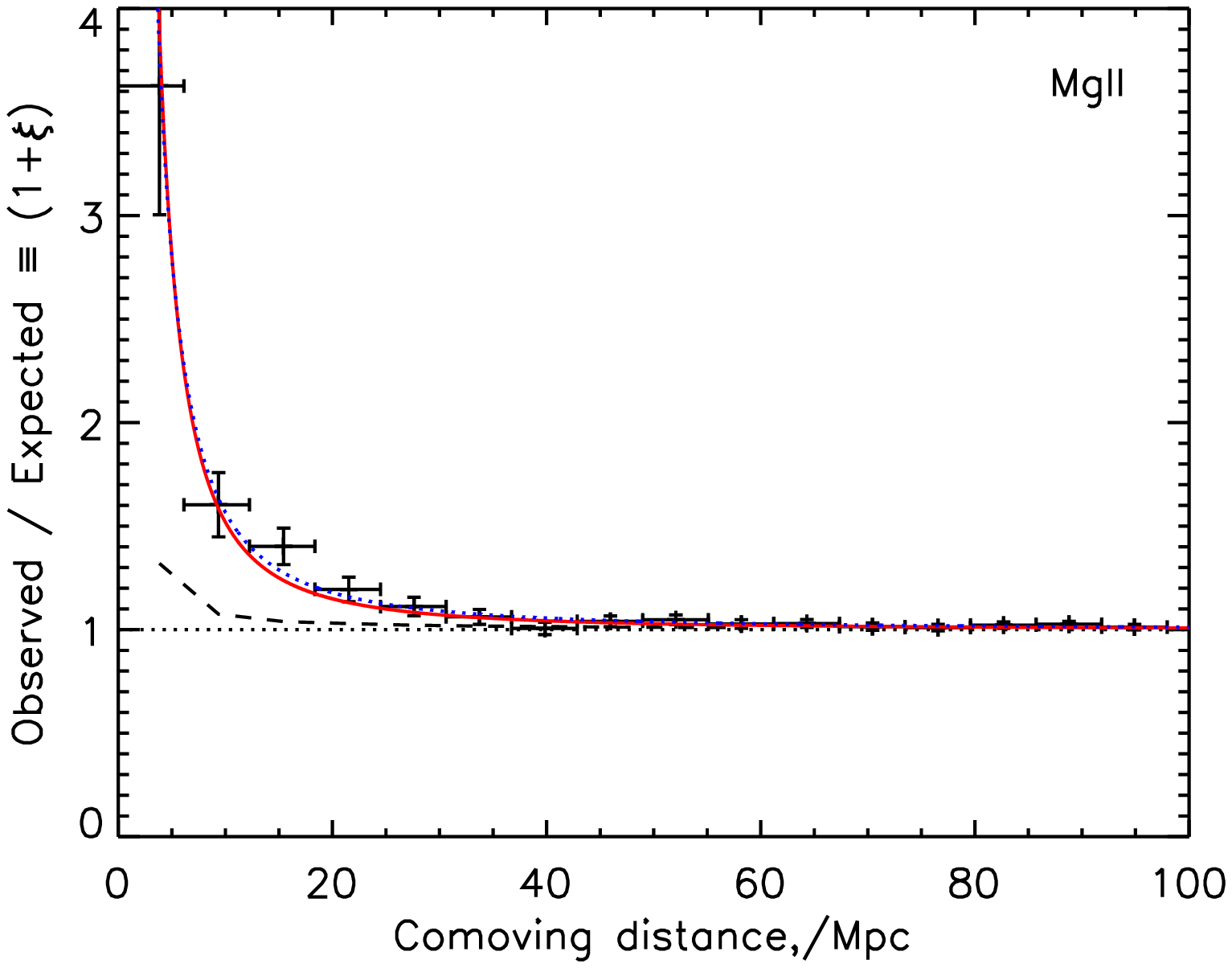}
\includegraphics[scale=0.5]{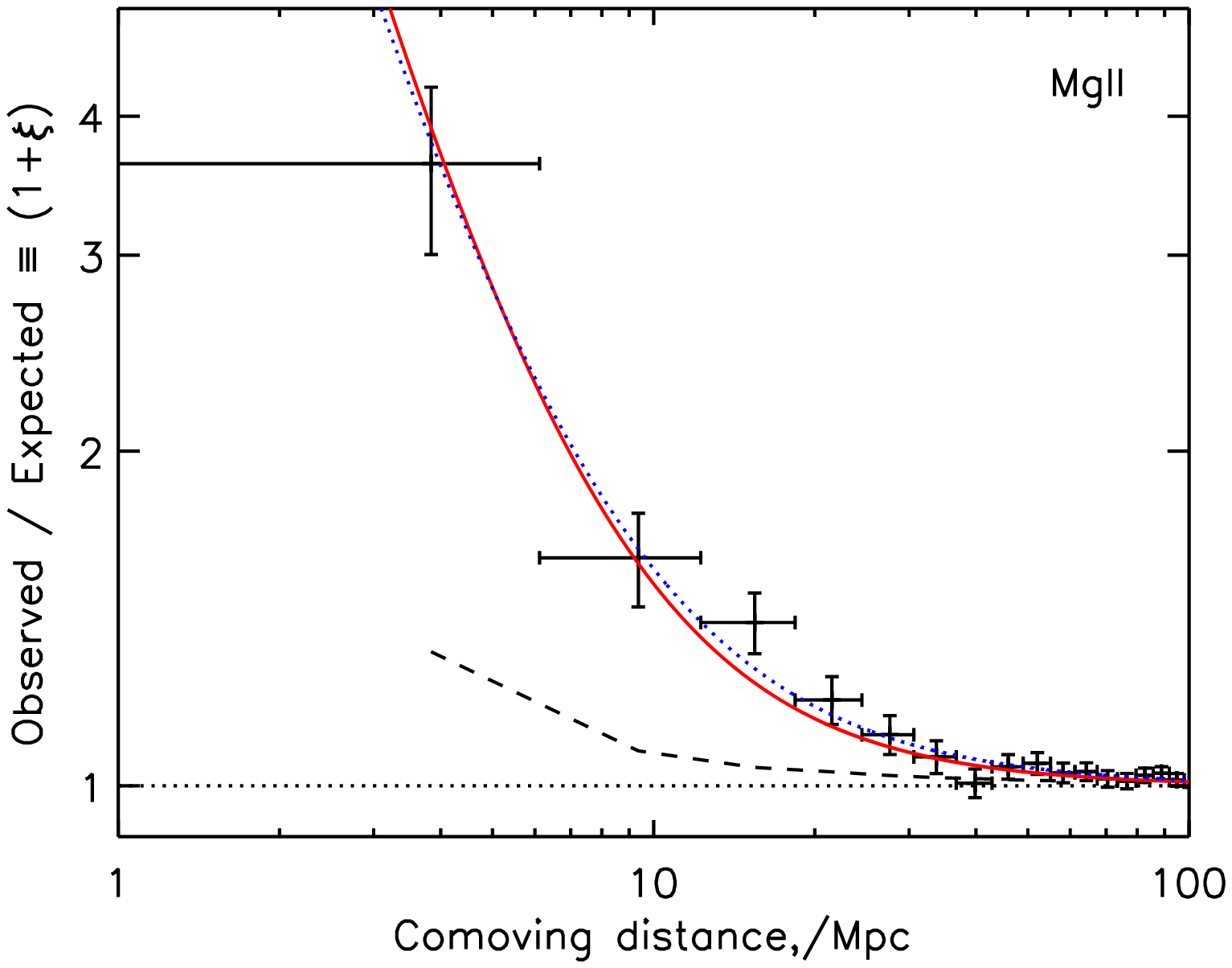}\\
\caption{{\it Top:} The 3-D correlation between \civ\ absorbers and
  QSOs (i.e. observed over expected number of pairs), as a function of
  comoving distance from the QSO. The right-hand panel is a
  logarithmic rendering of the left-hand panel. The overplotted plain
  line is the powerlaw described in the text, with fitted
  $r_0=5.8\pm1.1h^{-1}$\,Mpc for fixed $\gamma=1.8$. The dashed line
  is the 68\% detection threshold given the number of sightlines and
  number density of absorbers (Eq. \ref{eq:binomial}). {\it Bottom:}
  Same for \mgii\ absorber-QSO pairs. The additional dotted line is
  the fully fitted powerlaw, with  $r_0=5.0\pm0.4h^{-1}$\,Mpc and
  $\gamma=1.69\pm0.09$.  }\label{fig:trans}
\end{figure*}

The correlation functions for \civ\ and \mgii\ absorber-QSO pairs are
shown in Figure \ref{fig:trans}, where we plot ($1+\xi$) in bins of
$\sim6$\,Mpc. The results are plotted logarithmically in the right
hand panels. As expected, there is an enhancement in the counts of
both QSO-\civ\ pairs and QSO-\mgii\ pairs at small comoving distances from
the QSO. The significance of the detection is considerably higher for
the \mgii\ absorbers, because the sample is much larger.  Overplotted
as a dashed line is the 68\% confidence limit for detection of a
clustering signal, given the observed  and expected number of
absorbers in each bin (Eq. \ref{eq:binomial}).  In other
words, if there were no clustering, we would expect an independent bin
to lie above this line 32\% of the time.

We can evaluate the amplitude and significance of the clustering
signal for the \civ\ systems as follows.  The total fractional
excess\footnote{Fractional excess is defined as (N$_{\rm o}$-N$_{\rm
e}$)/N$_{\rm e}\equiv \xi$.} of \civ\ absorber-QSO pairs between
$\sim$6 and $\sim$43\,Mpc (bins 2--6) is 1.49. From
Eq. \ref{eq:binomial}, the probability of obtaining a value greater
than or equal to this value is 0.006 i.e. we detect excess clustering
at a 99.4\% confidence level.  If we instead use only the second to
fifth bins, we obtain a similar confidence level of 99.7\%. Including
the first bin results in a similar significance detection, but a lower
clustering amplitude. We have chosen to discard this first bin, which
only contains one pair, because of the uncertainty in the QSO
redshifts for this sample. Redshift errors will act to move clustered
QSO-absorber pairs to outer bins, and cause a flattening in the
correlation function at small distances. Our choice to discard the
first bin results in a small overestimation of the clustering
amplitude, as some QSO-absorber pairs from the central bin will be
counted in outer bins. The number of QSO-absorber pairs expected at
these small radii is very low, however, and their contribution to the
derived clustering amplitude is not significant within the final
errors.

For the \mgii\ absorber sample, the fractional excess within 43\,Mpc
is 4.0 and the detection of clustering is significant at the
$>$99.9999\% level.

These fractional excess numbers of QSO-absorber pairs (1.49 for \civ\ and
4.0 for \mgii) can be fit to the standard power law form used to
describe the correlation function, $\xi$ (Equation
\ref{eq:corrnfn}). To be consistent with typical values quoted in the
literature, we calculate $r_0$ for a fixed $\gamma=1.8$. We find $r_0
= 5.8\pm1.1$ and $4.84\pm0.4h^{-1}$\,Mpc for the \civ\ and \mgii\
absorbers respectively. These fitted power laws are overplotted in
Fig. \ref{fig:trans}. For the \mgii\ absorbers we can also fit the
correlation function to all bins directly, leaving $\gamma$ as a
free parameter. We use a non-linear least squares technique to obtain
$\gamma=1.67\pm0.09$ and $r_0=5.0\pm0.4h^{-1}$\,Mpc. The resulting
model has a reduced $\chi^2$ of 1.007. The curve is overplotted as a
blue dotted line in Fig. \ref{fig:trans}. In conclusion, in this
section we have unambiguously detected and measured the excess of
absorbers in the vicinity of QSOs due to the clustering of galaxies.

\section{A model for line-of-sight NALs}\label{sec:model}

In this section, we  model the line-of-sight correlation
of QSO-absorber pairs and we  address two questions: (i) is the central
spike at $\zabs\sim\zqso$ consistent with being caused by galaxy
clustering?  (ii) can the sub-relativistic tail of \civ\ absorbers be
caused by galaxy clustering? 

We proceed by constructing a model with three components: (i) the
amplitude of the clustering of absorption line systems arising from
galaxies in the vicinity of the QSOs; (ii) the velocity offset and
dispersion induced by peculiar motions and redshift errors; (iii) the
distance to which the QSO ionises surrounding gas.

Motivated by the results presented in the previous section,              
we construct a simple toy model of the form:
\begin{equation}\label{eq:model}
n_{abs}(r) = 
\begin{Bmatrix}
\left[1+\left(\frac{r}{r_0}\right)^{-1.8}\right]n_{bg} &r>R_{\rm cut} \\
0 & r\le R_{\rm cut} 
\end{Bmatrix}
*g(r,\sigma)
\end{equation}
where $n_{abs}$ is the mean number of absorbers per comoving Mpc at
comoving distance $r$ from the QSO, $n_{bg}$ is the
comoving number density of intervening absorbers,
$R_{\rm cut}$ is the comoving radius out to which the QSO
photoevaporates the absorbers (the ``proximity'' zone), and $r_0$ is the
correlation length of QSO-absorber clustering. 
The model is convolved with a Gaussian of width $\sigma$ to reproduce
the effects of the departure of the QSOs and absorbers from the Hubble
flow, as well as  errors on the redshift measurements of the QSOs.

Of the four parameters, $r_0$ is measured from the 3-D correlation
estimate (Section \ref{sec:trans}) and $n_{bg}$ is measured at
distances greater than 200\,Mpc from the QSOs. There are two remaining
parameters: $\sigma$ and $R_{\rm cut}$.  We will discuss plausible
values for these parameters in the following subsections. Our model is
no doubt a simplification of the true situation. For example, we do
not explictly account for a contribution from the halo clouds around
the QSO, assuming only a continuation of the powerlaw component
towards the QSO. In Section \ref{sec:results} we will discuss the
limitations of the model.

\subsection{Velocity dispersion}\label{sec:veldisp}

There are two contributions to the dispersion $\sigma$: redshift
errors on the QSO and, assuming the QSOs and absorbers originate in
different galaxies, peculiar motions of the absorbers and QSO host
galaxies with respect to the Hubble flow.

The SDSS catalogue provides errors on the redshifts of each QSO.  The
median quoted error of all the QSOs with absorption line systems are
0.0017 for the \mgii\ absorber sample and 0.0019 for the \civ\
absorber sample. This corresponds to velocity errors of 180\,km/s for
both samples at the median QSO redshifts (1.89 and 2.19 for the \mgii\
and \civ\ samples). However, given the observed systematic shifting of
broad QSO lines from the systemic redshift
\citep[e.g.][]{1982ApJ...263...79G,2000ARA&A..38..521S}, the true
errors are likely to be considerably larger, particularly for the
higher redshift \civ\ sample. As shown by \citet{2002AJ....124....1R}
\civ\ is shifted by a median $\sim$800\,km/s from the lower ionisation
\mgii\ line. The shift can be as much as 3000\,km/s in some QSOs. The
position of the \civ\ line is used in the template to derive the
systemic redshift of the SDSS QSOs and we may therefore expect this
shifting of the lines to cause errors of up to  $\sim$1000\,km/s.

The second contribution to the dispersion $\sigma$ comes from the
peculiar motions of galaxies with respect to the Hubble flow, if the
absorber and the QSO reside in different hosts. The pairwise velocity
dispersion of galaxies ($\sigma_{12}$) has been measured at low
redshift in the 2dFGRS and SDSS surveys
\citep{2004ApJ...617..782J,2006MNRAS.368...37L}. It can range from a
few hundred to nearly a thousand km/s, depending on the type of galaxy
\citep{2006MNRAS.368...37L}. However, the redshift evolution of
$\sigma_{12}$ has not been measured, making it difficult to predict
the velocity dispersion experienced by galaxies in the relatively
massive haloes of the QSOs. 

As neither the redshift errors nor the expected peculiar velocities of the
galaxies are well known, we choose to estimate $\sigma$ directly from
the data, by ensuring that our models reproduce the observed width of
the central peak in the comoving separation distribution (Figure
\ref{fig:dz}). The derived scatter in comoving distance is 9.0\,Mpc
for the \mgii\ absorber sample and 12.2\,Mpc for the \civ\ absorber
sample. This corresponds to velocity dispersions of 580\,km/s and
1030\,km/s at the median redshifts of the \mgii\
and \civ\ absorbers in the central peak of the comoving separation
distribution (1.35 and 1.97, see Section \ref{sec:results}).

It is important to note that the precise value chosen for $\sigma$
does not greatly affect our results; it only slightly alters the shape
of the central peak in the distribution of absorbers in velocity
space.

\subsection{The line-of-sight proximity effect}\label{sec:halo}

Two further physical scales are important for the geometry of
line-of-sight QSO-absorber clustering. The first is the distance to
which the gaseous halo of the QSO itself can contribute to the number of
pairs. The second is the distance out to which the intense radiation
field of the QSO eliminates the
particular ion in question. Here we simply present some typical scales
culled from the literature to guide in the interpretation of our
results,  $R_{\rm cut}$ is left as a free parameter in our model.

As discussed in the introduction, halos of \mgii\ around galaxies have
been studied for many years. Their proper extent is believed to be
$\sim$40$h^{-1}$\,kpc for unity covering fraction
\citep{1993eeg..conf..263S}. Beyond this distance absorbing clouds
have been found out to proper radii of $\sim100$\,kpc, but with covering factors
that decrease with distance \citep{2006ApJ...645L.105B,
2007ApJ...658..161Z, 2007arXiv0710.5765K}. Fewer quantitative studies
have been made for \civ, but \citet{2001ApJ...556..158C} find \civ\
halos of $\sim100h^{-1}$\,kpc in proper radius, with unity covering
factor. \civ\ and \mgii\ are believed to originate in the same halo
clouds, with the higher ionisation ion \civ\ almost always found in
conjunction with \mgii, but not the converse
\citep{1984A&A...133..374B,1992ApJS...80....1S,1999ApJ...519L..43C}.
The true extent of absorber clouds depends sensitively on the
equivalent width limit imposed,  also possibly on the mass of the host
halo \citep{1994ApJ...437L..75S,2008arXiv0801.2169C} and it is likely
that it changes with redshift. We therefore take these nominal values
for convenience, but the reader should keep in mind that they are only
indicative. 

The effect of the ionising field of the QSO on the surrounding
inter-galactic medium is less well constrained observationally
\citep[see e.g.][]{1994AJ....107.1219E}.  \citet{2007arXiv0706.4336C}
have used observational constraints from \mgii\ absorbers to model the
effect using the ionisation code CLOUDY
\citep{1998PASP..110..761F}. Assuming a uniform radiation field and a
cloud density $n=0.02{\rm cm^{-3}}$, they conclude that \mgii\ will be
completely destroyed in the inner 100\,kpc of the halo and that the
density of \mgii\ absorbers will only recover to the background level
on Mpc scales. They also predict that the fraction of \civ\ systems
will fall off rapidly below $\sim$200\,kpc. If the QSO radiation field
is anisotropic, the line-of-sight ionisation boundary must move
further out to remain consistent with
observation. \citet{2007ApJ...655..735H} also present a simple model
of the proximity effect for Lyman-limit systems. As \mgii\ systems are
found to correspond closely to Lyman-limit systems
\citep{1986A&A...169....1B,1992ApJS...80....1S}, these results are
relevant to our \mgii\ sample. They find absorbers with neutral
hydrogen column densities below $10^{19}$atoms/cm$^2$ and hydrogen
volume densities below $0.1{\rm cm^{-3}}$ to be photo-evaporated
within 1\,Mpc of the QSO.

\section{Results}\label{sec:results}

In Table \ref{table:1} we present the fraction of the QSOs in our
sample that have absorbers with $-0.01 < \beta < 0.01$ (associated)
and with $0.01 < \beta < 0.04$ (distant).  3.4\% (15\%) of QSOs in our
samples have one or more associated \mgii\ (\civ) absorbers with rest
equivalent width $>0.5$\AA. We note that these samples are at
different mean redshifts, and therefore the comparison of the
fractions must be treated with caution.  \citet{2003ApJ...599..116V}
find 25\% of QSOs to have \civ\ with the same EW limit within
$\pm$5000\,km/s. We find 18\% within this higher velocity limit, which
is consistent given the small number of objects in the
\citet{2003ApJ...599..116V} sample.  Given the fact that \mgii\ and
\civ\ are almost always seen if a QSO sightline passes close to a
galaxy, these results indicate that the QSO must ionise most of the
\mgii\ and \civ\ clouds in its own halo.

In this section it is often necessary to convert between physical and
comoving scales. As we are primarily interested in describing the
central peak of the comoving separation distribution, we use the
median redshift of those QSOs contributing to the peak. These are 1.35
and 1.97 for the \mgii\ and \civ\ samples respectively.

\begin{table*}
\begin{center}
\caption{The fraction of QSOs with one or more strong \mgii\ or \civ\
   absorption lines within fixed velocity ranges. The absorber rest
   frame EQW limit is 0.5\AA\ and the radio-quiet (RQ) and radio-loud (RL) samples have been
   matched in K-corrected $i$-band absolute magnitude. }\label{table:1}
\begin{tabular}{c|c|c|c|c|c|c}
velocity &  F$_{\rm MgII(all)}$ & F$_{\rm MgII(RQ)}$ & F$_{\rm MgII(RL)}$
& F$_{\rm CIV(all)}$ & F$_{\rm CIV(RQ)}$ & F$_{\rm CIV(RL)}$ \\\hline\hline
$-0.01<\beta < 0.01$   & 0.034 & 0.036 & 0.054 & 0.15 & 0.14 & 0.16 \\     
$0.01<\beta<0.04$ & 0.044 & 0.051 & 0.055 & 0.11 & 0.11 & 0.12  \\    \hline
\end{tabular}
\end{center}
\end{table*}
   
\subsection{The $\Dz\sim0$ excess}  

\begin{figure*}
\includegraphics[scale=0.5]{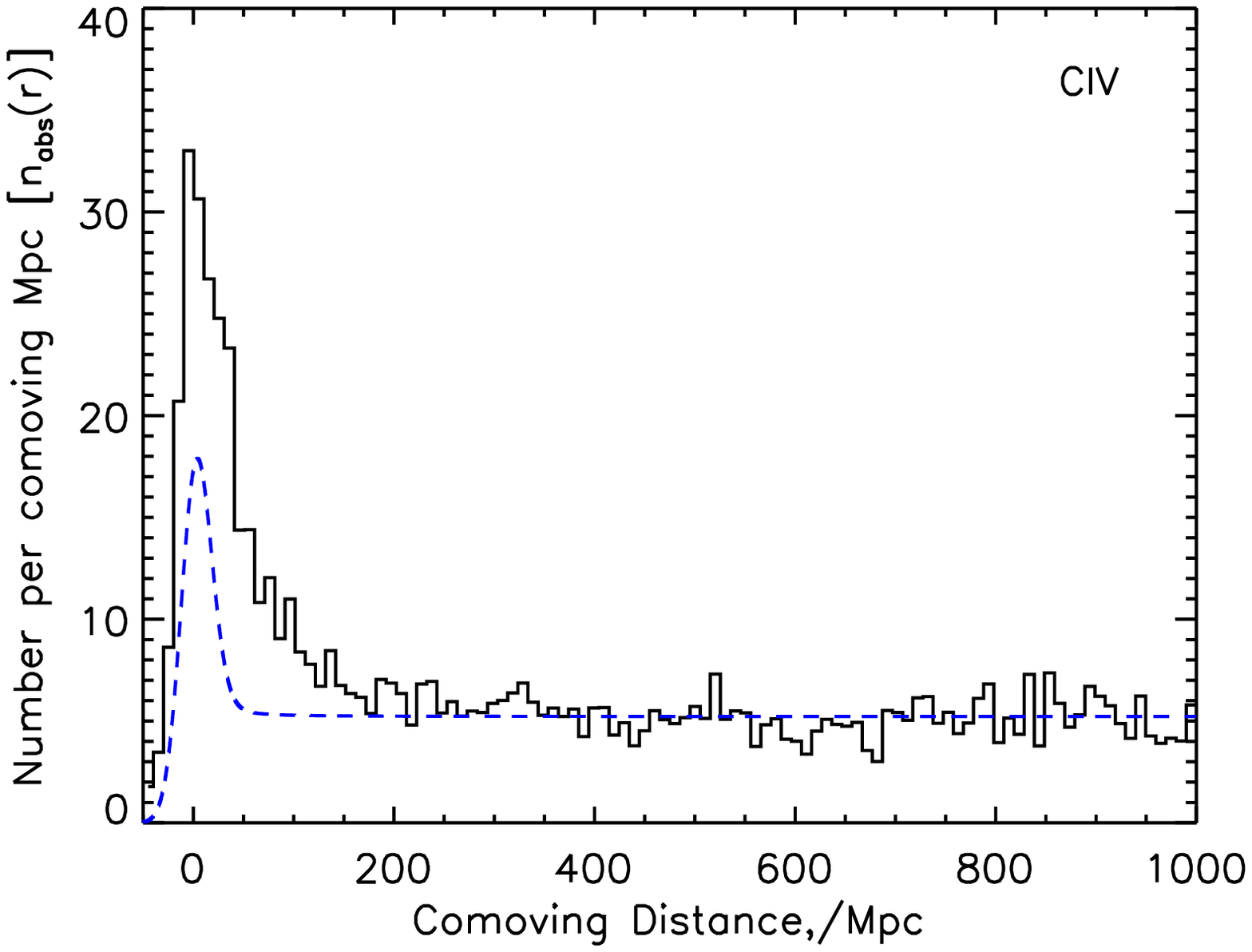}
\includegraphics[scale=0.5]{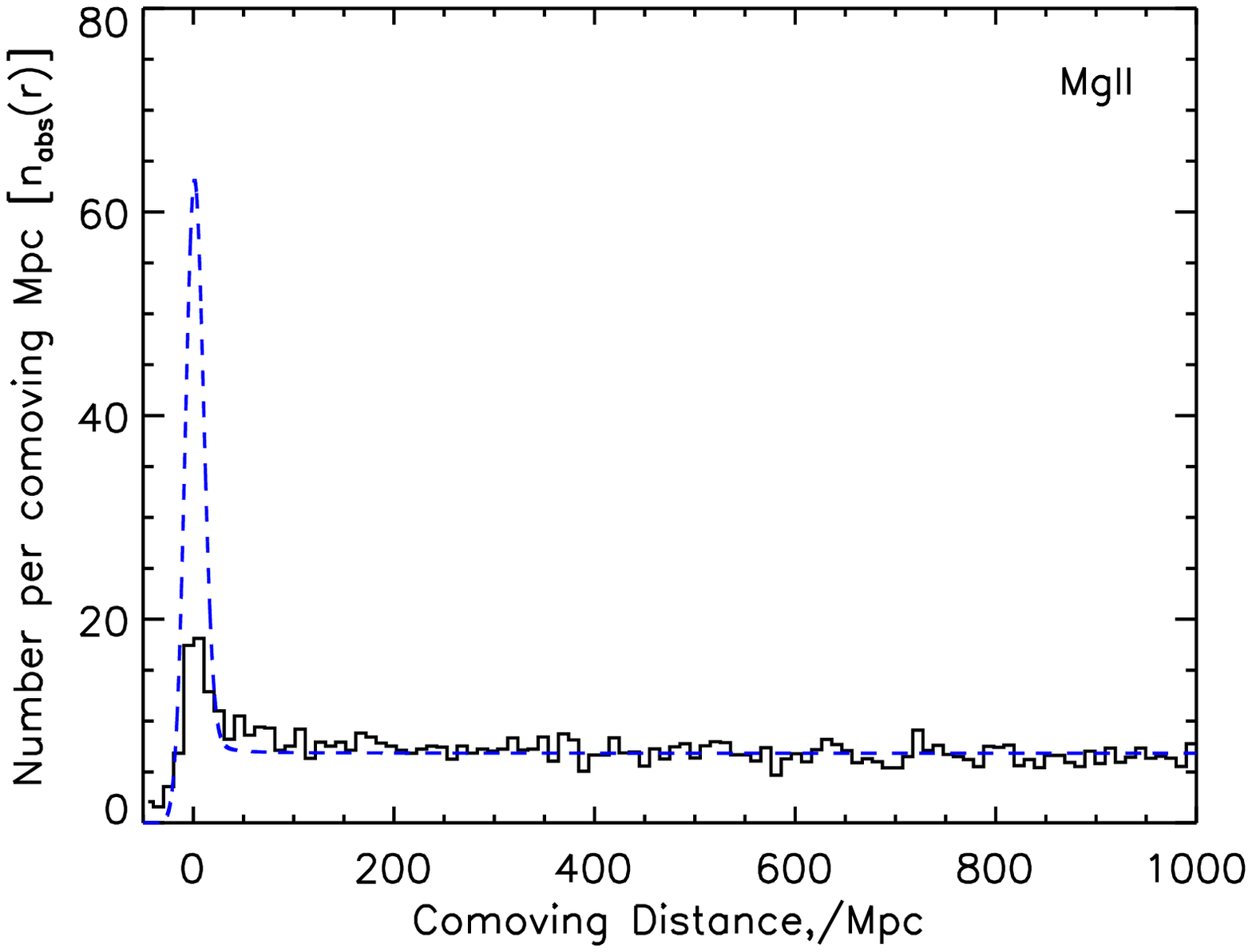}
\caption{{\it Left}: The number of \civ\ absorbers per comoving Mpc as
  a function of line-of-sight comoving distance from the
  QSO. Overplotted as a dashed line is the modelled galaxy clustering
  component, assuming the QSO photoevaporates a sphere corresponding
  to previous estimates of the typical size of absorber halos. {\it
  Right}: same for \mgii\ absorbers. }\label{fig:los1}
\end{figure*}

We now compare the line-of-sight distributions of the NALs with the
prediction of our model for the component due to galaxy clustering.
We first assume that the QSO ionises clouds only in its own halo,
i.e. we set $R_{\rm cut}=R_{\rm halo}$ in Equation \ref{eq:model}. We
take $R_{\rm halo}$ to be the comoving size of absorber halos at the
median redshifts given above assuming typical sizes from the
literature, i.e. 130 and 420\,kpc for the \mgii\ and \civ\ samples
respectively. The predicted line-of-sight absorber distributions are
shown in Figure \ref{fig:los1}. There are two clear results, each with
an obvious implication. Firstly, galaxy clustering only contributes to
the central $\sim50$\,Mpc of the distribution. We therefore conclude
that the tail of high-velocity absorbers seen in the \civ\
distribution must be intrinsic to the QSO and/or its host
galaxy. Secondly, the model greatly overpredicts the number of
\mgii\ absorbers, leading us to conclude that the QSOs ionise \mgii\
clouds well beyond the halo of the host galaxy. The model
overpredicts the \mgii\ counts within $\pm$40\,Mpc of the QSO
(corresponding to $\pm$2500\,km/s) by a factor of 2.05.

\begin{figure*}
\includegraphics[scale=0.5]{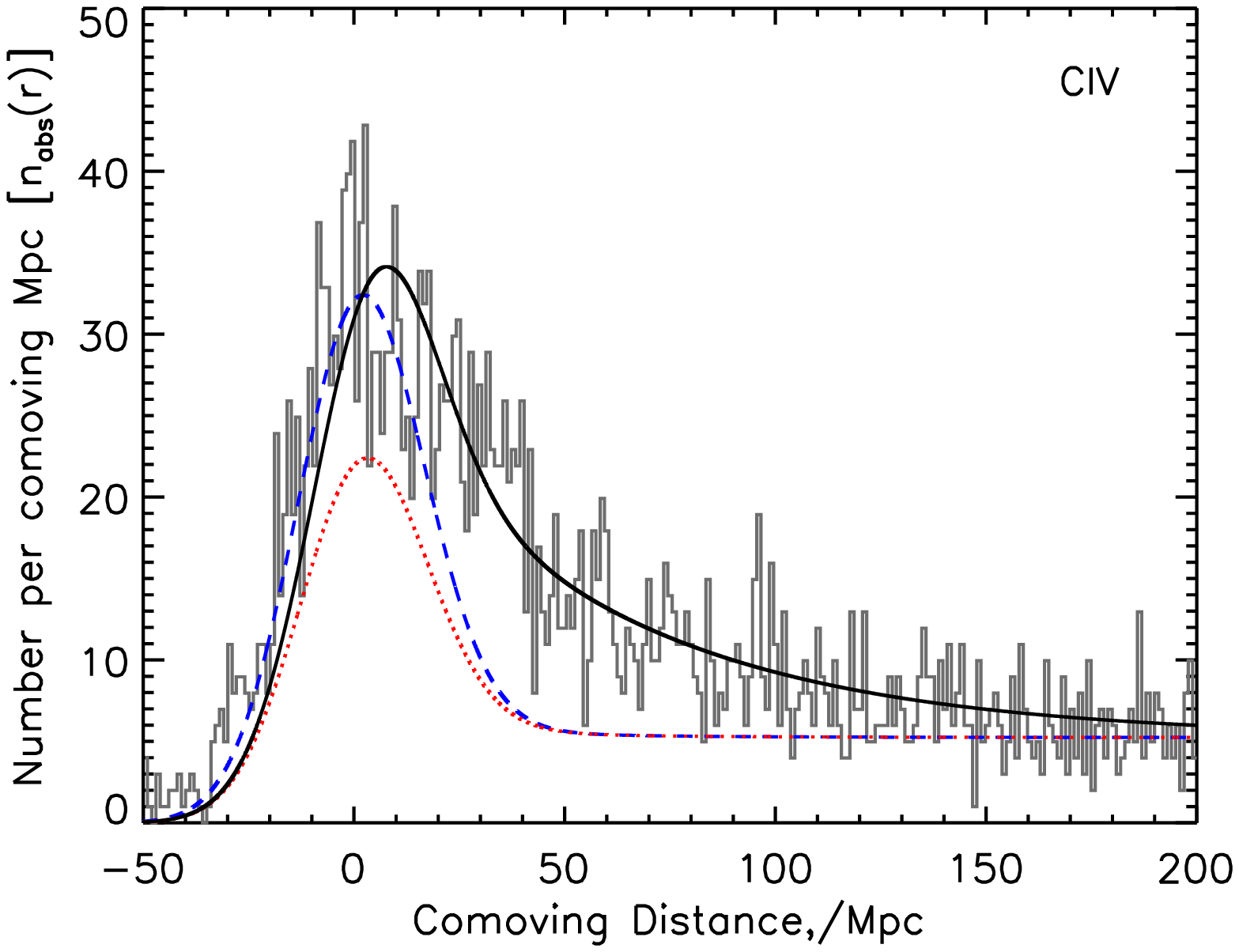}
\includegraphics[scale=0.5]{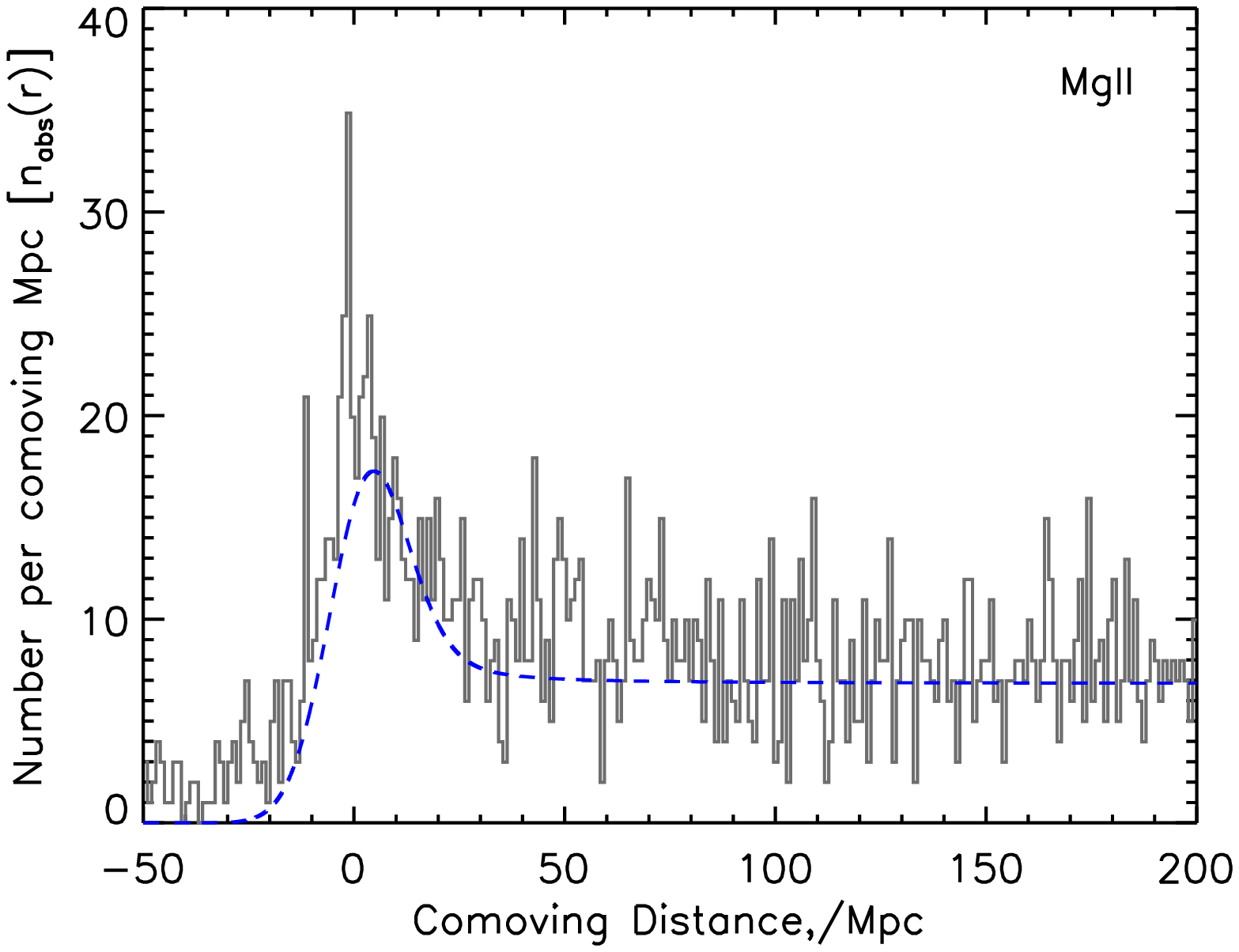}
\caption{{\it Left}: The number of \civ\ absorbers per comoving Mpc as
  a function of line-of-sight comoving distance from the QSO,
  focussing on the region close to the QSO. Overplotted as a dashed (blue)
  line is the modelled galaxy clustering component, assuming the QSO
  photoevaporates a sphere of radius 0.18\,Mpc, smaller than the
  observed size of \civ\ halos at low redshift.  The full line is for
  the full model with $R_{cut}=0.3$\,Mpc and including the intrinsic,
  outflowing absorber distribution. The dotted (red) line is the
  clustering component of this final model. {\it Right}: same for
  \mgii\ absorbers, with the clustering model assuming a proximity sphere of
  radius 0.8\,Mpc, more than six times larger than the \mgii\ haloes
  observed at low redshift, and assuming no intrinsic component. }\label{fig:los2}
\end{figure*}

By increasing the size of the region over which the QSO ionises \mgii\
($R_{\rm cut}$), our simple model can reproduce the observed distribution
of \mgii\ absorbers reasonably well (albeit with a small offset that
we will address below).  Figure \ref{fig:los2} shows the line-of-sight
distribution of \mgii\ and \civ\ absorbers within 200\,Mpc from the
QSO. Overplotted in the right hand panel is a model with $R_{\rm
cut}=0.8$\,Mpc (close to the size of the proximity zone for
Lyman-limit systems derived by Hennawi \& Prochaska [2007]).

Although our model with a large proximity zone appears to fit the
\mgii\ data rather well, we note that our simple "top-hat" exclusion
zone is no doubt an over-simplification. Some \mgii\ absorbing clouds
probably do exist within 0.8\,Mpc of the QSO. Such systems may be
associated with material within QSO host galaxies.  Vanden Berk et~al
(2008) have shown that such associated \mgii\ systems tend to be on
average dustier (implying denser gas) and be more highly ionised (as
would be expected for clouds within $R_{\rm cut}$). By increasing the
radius out to which \mgii\ halo clouds of average volume density are
evaporated, our model would allow for an increasing fraction of the
associated absorbers to be directly linked to the presence of the QSO
and its host galaxy. We will return to this point in Section
\ref{sec:highv}.

Unlike the \mgii\ systems, the \civ\ absorber distribution does not
require a large proximity zone. Figure \ref{fig:los1} shows that if
the QSO photoevaporates all \civ\ clouds in its own halo, the number
of absorbers within $\pm$40\,Mpc is {\em underpredicted} by more than
a factor of two.  We can try to fit the distribution by decreasing
$R_{\rm cut}$; the model over-plotted as a dotted line in the
left-hand panel of Figure \ref{fig:los2} has $R_{\rm cut}=0.18$\,Mpc,
corresponding to a proper radius of 60\,kpc at the median redshift of
the QSOs. This radius is smaller than the estimated, but uncertain,
\civ\ halo size discussed in Section \ref{sec:los}.  We note that such
a small proximity radius may conflict with the small fraction of QSOs
observed to have associated \civ\ absorption (16\%). We will return to
this point in the following sections.

The main conclusions of this section as follows:                         
\begin {enumerate}
\item the central peak of the \mgii\ absorber distribution could be
explained by galaxy clustering if the QSO ionises absorbers to
$\sim0.8$\,Mpc. Assuming the physical size of \mgii\ halos remains the
same as observed at low redshift, this corresponds to more than 6
times the size of the QSOs own \mgii\ halo, implying that absorbers in
the halos of neighbouring galaxies along the line-of-sight are also
affected. 
\item The ionising sphere must be substantially larger for the \mgii\
absorption line systems than for the \civ\ absorption line systems,
consistent with the higher ionisation energy of the latter ion. 
\item The \civ\ systems exhibit a high velocity tail of absorbers,
which cannot be explained by galaxy clustering and hence must be {\em
intrinsic}.  It is likely that these intrinsic absorbers also
contribute to the distribution at lower velocities.
\end{enumerate}

In the following subsection we will further investigate the
distribution of these intrinsic absorbers.

\subsection{The high velocity \civ\ absorbers}\label{sec:highv}

From Figure \ref{fig:los2} it is clear that many narrow \civ\
absorbers are in a high velocity tail that extends out to 
150\,Mpc before becoming indistinguishable from the background
absorber population. 
This tail cannot be due  to galaxy clustering, and we conclude that
these absorbers are truly intrinsic to the QSO.

\begin{figure*}
\includegraphics[scale=0.5]{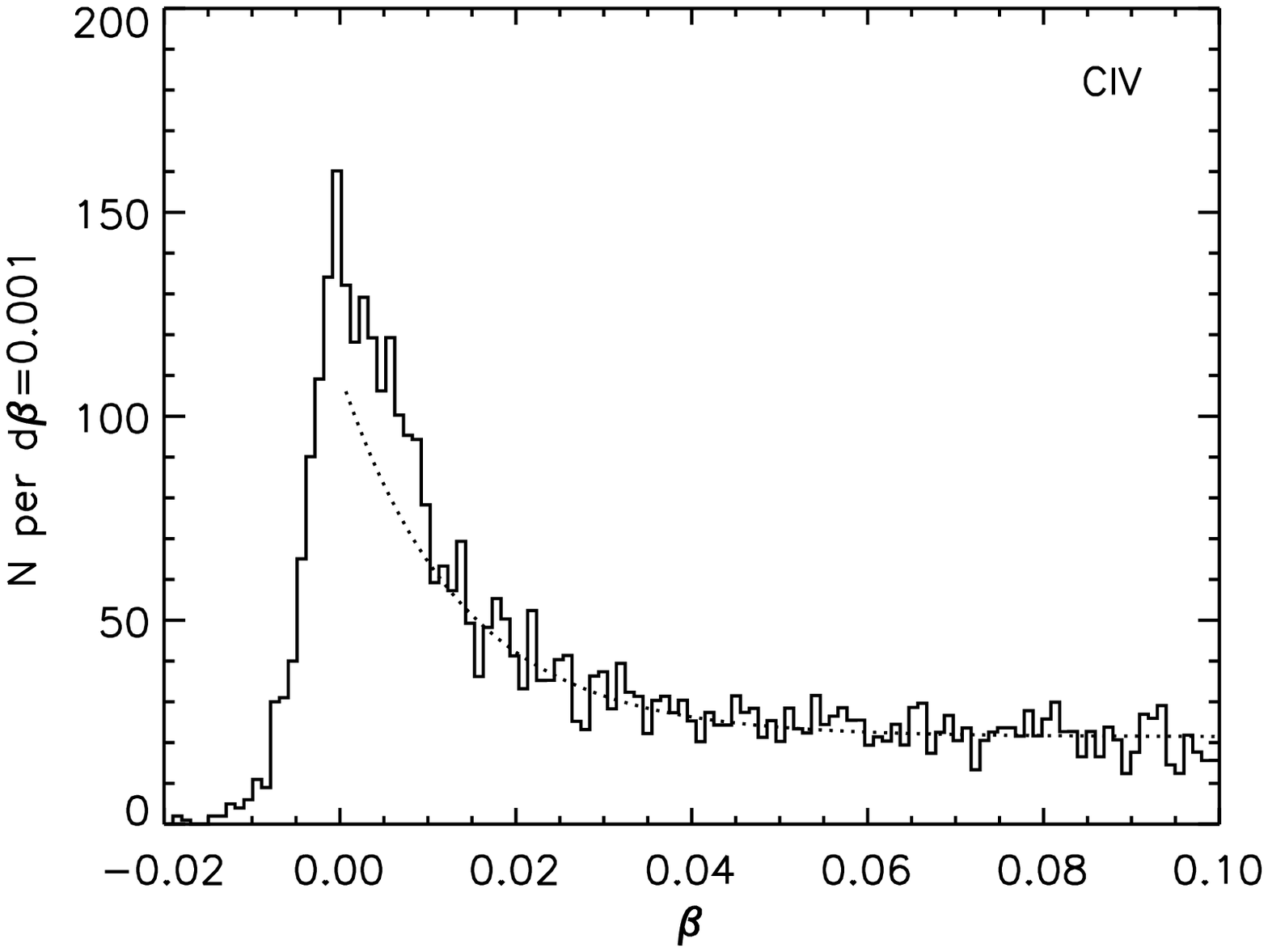}
\includegraphics[scale=0.5]{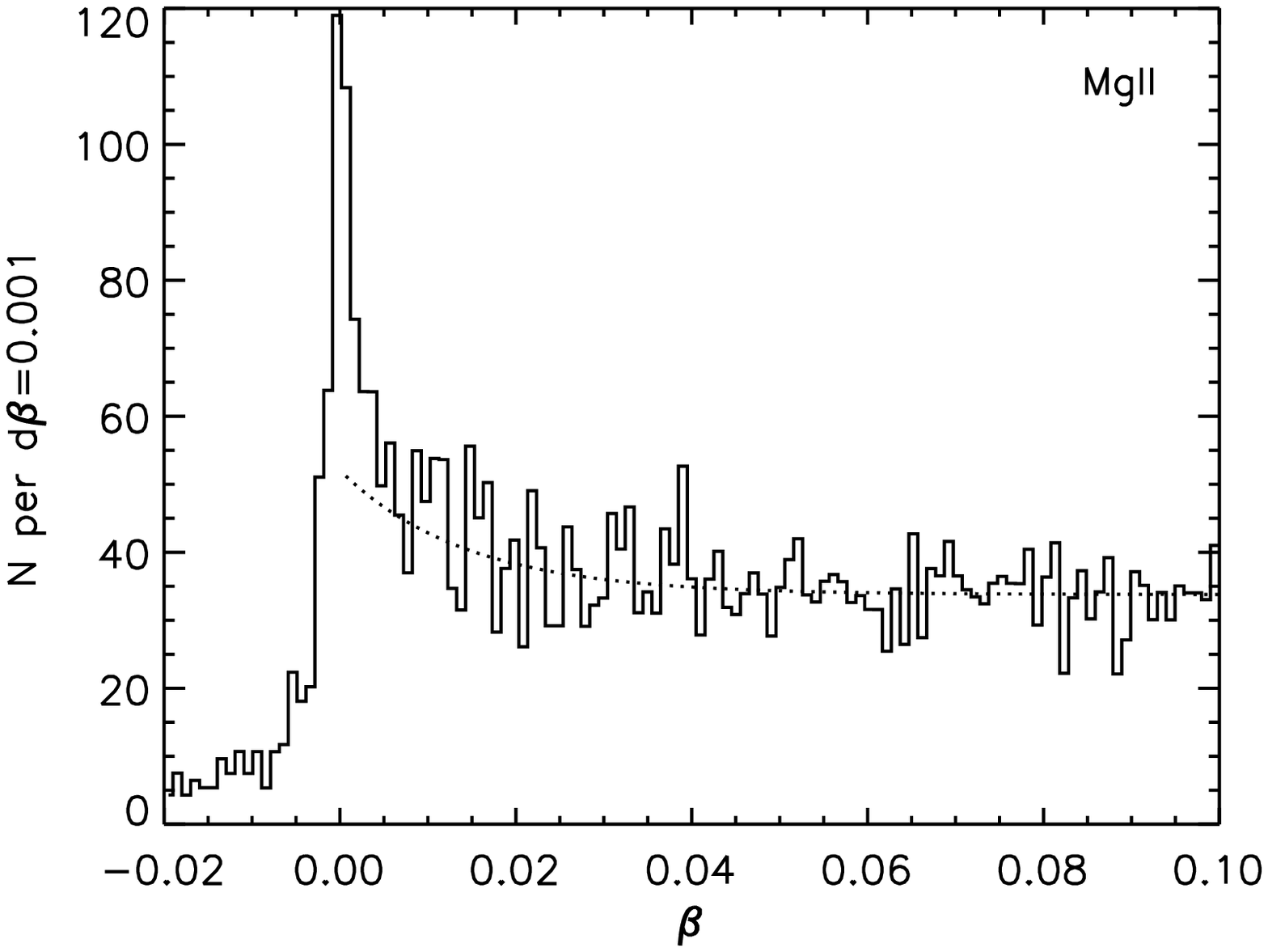}
\caption{The $\beta\equiv v/c$ distribution for \civ\ and
  \mgii. Overplotted as a dashed line on the \civ\ high-velocity tail
  is an exponential curve $e^{-a\beta}$ with $a=70$. A curve of the
  same width is fitted to the \mgii\ high-velocity tail, with a height
  that is 20$\pm$5\% that observed in the \civ\ absorbers. }\label{fig:beta}
\end{figure*}

In Figure \ref{fig:beta} we present a simple model for the velocity
distribution of the absorbers.  We have fit an exponential to the
high-velocity component:
\begin{equation}
n_{\rm abs} -n_{bg}  \propto e^{-a\beta} 
\end{equation}
where
\begin{equation}
\beta \equiv\frac{v}{c} = \frac{R^2-1}{R^2+1},
\end{equation}
\begin{equation}
R = \frac{1+z_{\rm qso}}{1+z_{\rm abs}},
\end{equation}
and $a=70\pm7.7$.  The model is fit in the range  $0.01<\beta<0.05$ to
avoid the central clustering peak. 

Our model predicts that between 3000 and 12000\,km/s
($0.01<\beta<0.04$), 45\% of \civ\ absorbers are caused by
sub-relativistic outflowing material from the QSO.  Due to the
presence of the clustering component, we cannot constrain the shape of
this intrinsic component at low-velocities. While it seems unlikely
that the number of intrinsic absorbers drops sharply towards low
velocities, various processes may act to enhance their number (see
Section \ref{sec:radio}). Our derived fraction of 45\% intrinsic
absorbers may thus be an underestimate.

Along similar lines, we note that a small excess of \mgii\ absorbers
with $0.005<\beta<0.015$ may be present in Figure \ref{fig:los2}. We
have attempted to constrain the strength of this excess by fitting an
exponential distribution with a width fixed to be that observed in the
\civ\ absorbers. We obtain a height which is 20$\pm$5\% that found for the
\civ\ absorbers.

We now attempt to build a model to describe both the extrinsic and
intrinsic \civ\ systems.  We convert our modelled velocity
distribution of outflowing material into a distribution in comoving
distance by assuming that each QSO contributes an equal proportion of
the high-velocity tail. For each absorber-QSO pair we transform the
model velocity distribution into the distribution in comoving distance
at the redshift of the QSO. We convolve the final distribution with a
Gaussian to account for the QSO redshift errors and add this to the
clustering model with different values of the parameter $R_{\rm
cut}$. The result with $R_{\rm cut}=0.3$ is shown in Figure
\ref{fig:los2} as the solid black line. The contribution from
clustering is shown by the dashed red line. The model fits well and it
implies that a substantial fraction of low-velocity \civ\ absorbers,
with $v<3000$\,km/s, may also be part of a distribution of outflowing
material that is intrinsic to the QSO and/or its host. This increased
value for $R_{\rm cut}$ is now about 70\% that of low redshift \civ\
halos. Together with the likelihood that $R_{\rm cut}$ varies with
redshift, QSO luminosity and halo mass, the fact that only 16\% of
QSOs show associated absorption does not appear incompatible with our
model.

In Table \ref{table:2} we present the fraction of each component
(background, galaxy clustering, intrinsic) present in our model in
different velocity ranges for both \mgii\ and \civ\ absorbers. The
number of absorbers associated with intervening galaxies in our model
is the sum of the background and galaxy clustering numbers. We note
that the fractions of intrinsic absorbers presented here may be lower
limits; we will return to discuss this in Section \ref{sec:disc}. 

\begin{table*}
\caption{The fraction of \mgii\ and \civ\ absorbers above the absorber
  rest frame EQW limit of 0.5\AA\ attributed to background (bg),
  galaxy clustering (gc) and intrinsic (int) components in our final
  models. A comoving distance of 40\,Mpc corresponds approximately to
  a $\beta$ of 0.008 (0.014) at the median redshift of the QSOs in the
  \mgii\ (\civ) central peak. A comoving distance of 170\,Mpc
  corresponds approximately to a $\beta$ of 0.037
  (0.07). }\label{table:2}
\begin{tabular}{c|c|c|c|c|c|c}
velocity &  F$_{\rm MgII(bg)}$ & F$_{\rm MgII(gc)}$ & F$_{\rm MgII(int)}$
& F$_{\rm CIV(bg)}$ & F$_{\rm CIV(gc)}$ & F$_{\rm CIV(int)}$ \\\hline\hline
$-40 <r/{\rm Mpc} < 40$   & 0.46 & 0.54 & 0.0 & 0.13 & 0.48 & 0.39\\     
$40<r/{\rm Mpc}<170$ & 0.99 & 0.01 & 0.0 & 0.53 & 0.01 & 0.45 \\    \hline
\end{tabular}
\end{table*}

\subsection{Radio Loud and Radio Quiet QSOs}\label{sec:radio}

A key point of interest is whether the line-of-sight distribution of
absorbers differs between radio-loud (RL) and radio-quiet (RQ) QSOs.
The SDSS QSO catalogue has been matched with the FIRST radio survey
\citep{1995ApJ...450..559B}. In this section we have included only
those QSOs in the FIRST survey footprint. RL QSOs have been defined to
be those with 1.4GHz luminosities greater than $10^{25}$W/Hz
\citep{1990MNRAS.244..207M}. The FIRST radio survey is flux limited to
1\,mJy, which corresponds to $\sim10^{25}$W/Hz at a redshift of 2,
close to the mean redshift of the \civ\ absorbers. We therefore expect
our RL sample to be complete in the \mgii\ sample, but some RL QSOs in
the \civ\ absorber sample will contaminate the RQ sample. As the
fraction of RL QSOs is low, around 10\% of all QSOs, this will not
significantly bias the RQ distribution.  In the \mgii\ sample we have
1298 RL and 13179 RQ QSOs.  In the \civ\ sample we find 543 RL and
5355 RQ QSOs.

\begin{figure*}
\includegraphics[scale=0.5]{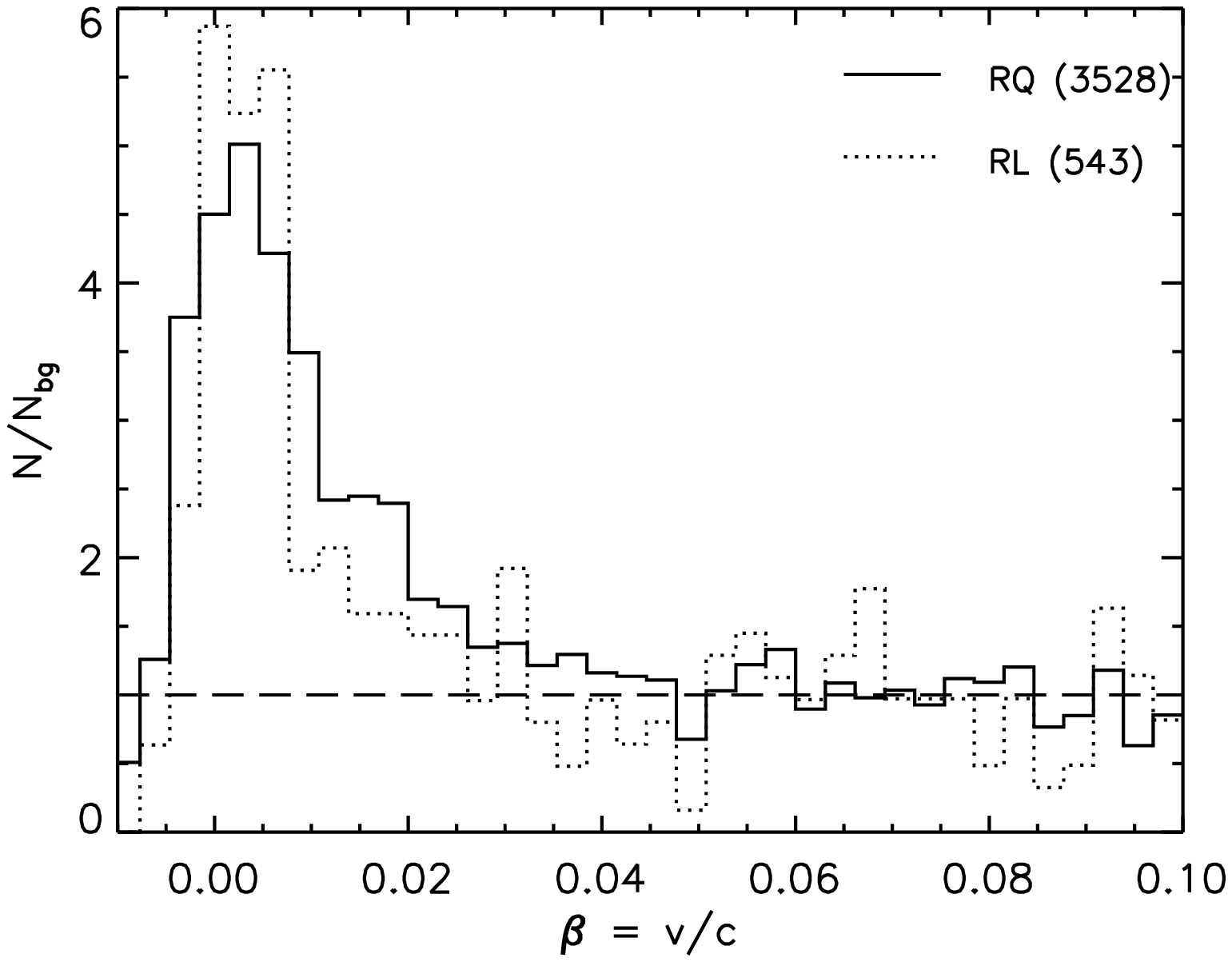}
\includegraphics[scale=0.5]{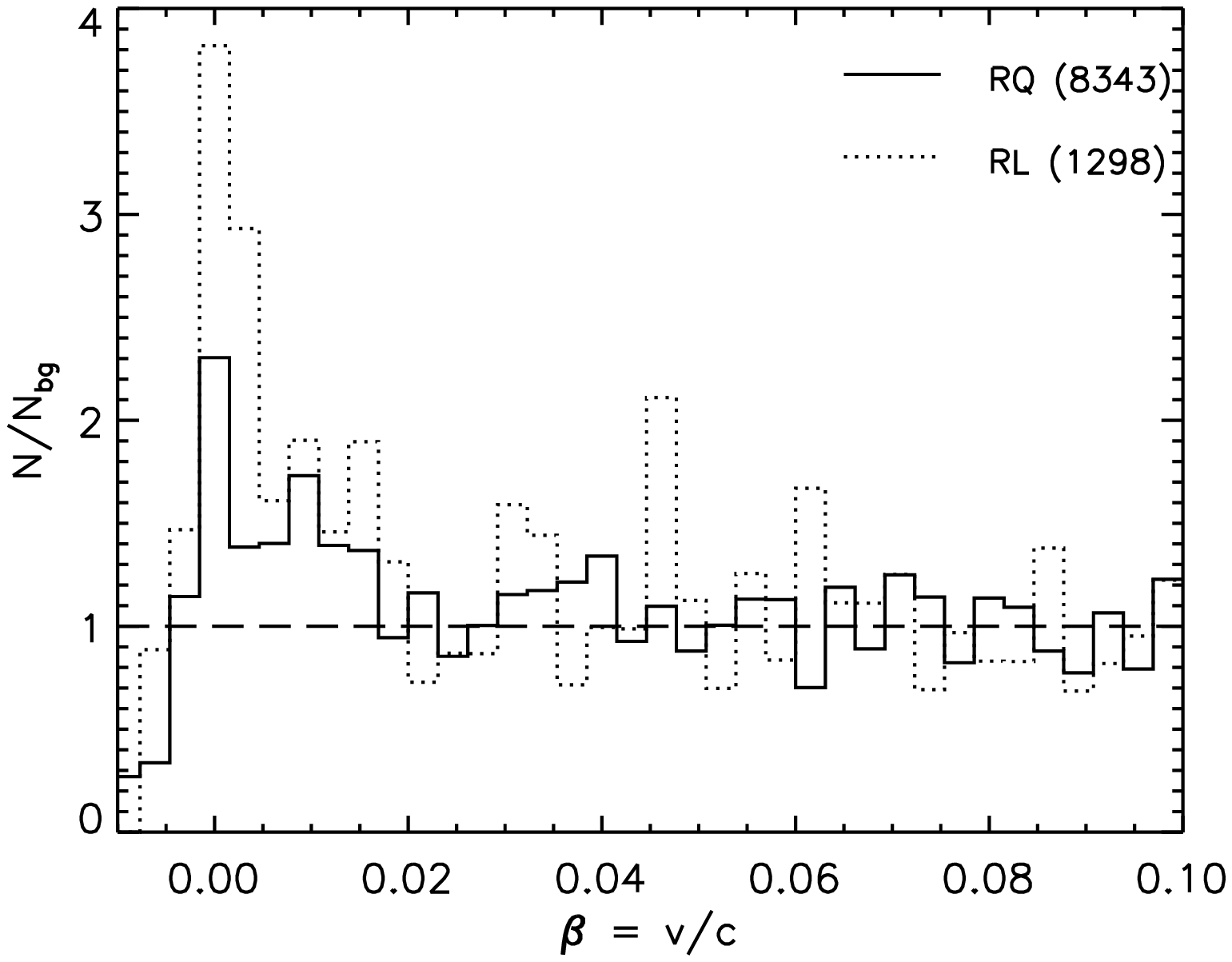}
\caption{{\it Left:} the fraction of \civ\ absorbers relative to the
  background level found in RQ (full line) and RL (dashed line) QSOs
  as a function of velocity shift from the QSO. The peaked excess
  close to the QSO is more pronounced in the RL QSOs but the
  relativistic tail of absorbers exists in both RQ and RL QSOs. The
  long-dashed line is positioned at unity for reference. The total
  number of QSOs in each sample is given in the top right. {\it Right:}
  Same for \mgii\ absorbers. }\label{fig:db_radio}
\end{figure*}

It is known that a correlation exists between optical and radio
luminosity \citep[e.g.][]{2003MNRAS.346..447C,2007ApJ...658..815S}.
We have therefore matched the optical luminosities of our RQ QSO
sample to those of our RL QSO sample to avoid any biases. Optical
luminosity is defined from the $i$-band PSF magnitude, K-corrected to
$z=2$ following \citet{2006AJ....131.2766R}. This matching procedure
reduces the RQ samples to 8343 and 3528 for the \civ\ and \mgii\
samples respectively. The velocity distributions are shown in Figure
\ref{fig:db_radio}.

Comparing the RL and RQ samples, we see clear differences. A narrower
and more pronounced spike at $\beta\sim0$ is seen for the RL QSOs for
both \civ\ and \mgii\ absorbers. This result for \civ\ has been
presented for many fewer QSOs by \citet{2003ApJ...599..116V}. We also
note that the subrelativistic tail of \civ\ absorbers is visible in
both the RL and RQ samples, but is clearly more pronounced in the RQ
QSOs. This high-velocity tail was observed previously in RQ QSOs by
\citet{1999ApJ...513..576R} and \citet{2001ApJS..133...53R}. We will
discuss these results in more detail in Section \ref{sec:disc_radio}.

\section{Discussion}\label{sec:disc}

Associated absorption line systems account for a small fraction of all
metal absorbers  detected in spectroscopic QSO surveys; however,
their study is important for understanding the effect of QSOs on their
host galaxies and their local environment. In the following subsections we
will address the implications of our results for understanding the
nature of associated narrow absorbers and for absorption line systems
in general.

\subsection{Intrinsic outflowing gas}

The primary aim of our paper is to quantify whether some narrow
absorption line systems are intrinsic to the QSO and its host
galaxy. From our analysis alone, there is clear evidence that a high
fraction of \civ\ absorbers are intrinsic to the QSO.  45\% of \civ\
systems with velocities in the range $0.01<\beta<0.04$ are intrinsic
and are well described by an exponential velocity distribution. A
similar, but considerably smaller fraction of high-velocity \mgii\
absorbers are attributable to the QSO. The high-velocity \mgii\ tail
is 20$\pm$5\% the height of that observed in \civ. Starburst-driven
winds do not reach such high velocities; it is therefore clear that
these absorption line systems are a direct consequence of a QSO
induced outflow \citep[see compilation in Figure 2
of][]{2007ApJ...663L..77T}.

The physical radius at which the intrinsic \civ\ NALs originate cannot
be constrained from these results alone. It is possible that they
originate from the wind close to the accretion disk of the central AGN
itself, although it may be difficult to understand the narrowness of
the lines under these circumstances.  A key constraint is that narrow
intrinsic absorbers are observed in both the RL and RQ QSOs, contrary
to BALs, the incidenc of which decreases sharply as radio power
increases
\citep{1992ApJ...396..487S,1997ApJ...479L..93B,2008arXiv0801.4379S}. One
possible explanation is that the strong X-ray flux observed in RL QSOs
limits the self-shielding of the winds and thus the widths of the
absorption features \citep{1995ApJ...451..498M}. In this scenario NALs
are simply failed BALs. Alternatively, NALs may be due to the viewing
angle of the accretion driven wind \citep{2000ApJ...545...63E}; the high velocities observed in
this work will allow tight constraints to be placed on the wind
geometry.

It is also possible that some of these absorbers, particularly those
with lower velocities, may not be directly associated with the
accretion disk winds, but originate further out in the ISM of the
galaxy.  The QSO spectra contain significantly more information than
we have used in the current paper: for example, it is possible to
derive their dust
content and ionisation states \citep[see][]{vdb_assabs} and also to look
for fine structure lines and line variability in multi-epoch
observations. Stacking the spectra as a function of velocity distance
from the QSO will increase the signal-to-noise of the weaker lines
needed to obtain accurate column densities
\citep{2006MNRAS.367..211W,2006MNRAS.tmp..274Y}. All of this
information should enable us to constrain more precisely the physical
origin of the \civ\ outflows, just as it has allowed \citet{vdb_assabs} to
suggest an origin for some associated \mgii\ systems in the gas of the
QSO host galaxy.

\subsection{The proximity effect of the QSO}

Our results show that the QSO has a considerable impact on NALs along
the line-of-sight, ionising ordinary \mgii\ halo clouds well beyond
the expected \mgii\ halo size of the QSO host galaxy. The QSO also
destroys a considerable fraction of the \civ\ clouds within the host
halo. Our model based on galaxy clustering overpredicts the number of
\mgii\ absorbers within 40\,Mpc of the QSO by a factor of two.  This
is somewhat less than the factor of 4-20 estimated by
\citet{2007ApJ...655..735H}, who studied the distribution of 17
Lyman-limit systems around QSOs.

The effective radius of the first bin in our 3 dimensional \mgii\
absorber-QSO cross- correlation analysis is 4.2\,Mpc.  Our results can
thus test whether there is a ``transverse proximity effect'' on this
scale \citep[e.g.][]{2007arXiv0711.4113G}. We fit the \mgii\
correlation function (Equation \ref{eq:xi}), this time excluding the
central bin. The resulting power-law predicts 38 pairs in the central
bin, compared to the 34 we actually observe. Thus, we find no evidence
for a deficit of \mgii\ absorber-QSO pairs at this radius. This is in
agreement with the results of \citet{2006ApJ...645L.105B},
\citet{2007ApJ...655..735H} and \citep{2007arXiv0711.2308T}, who find
no transverse proximity effect for \mgii, Lyman Limit systems, or for
metal-line systems in general.

\subsection{Radio loud vs. Radio quiet}\label{sec:disc_radio}

Unfortunately our sample of RL QSOs is too small to allow us to
measure the 3D correlation function of RLQSO-absorber pairs, so we can
not assess the origin of the larger excess of low-velocity absorbers
in RL QSOs, visible in both the \mgii\ and \civ\ distributions.

There are several possible explanations for the excess number of
low-velocity absorbers seen in the RL QSOs:
\begin{itemize}
\item the radio jets themselves drive low-velocity, lower ionisation
or clumpier outflows,
\item RLQSOs ionise their surrounding medium to smaller distances than
RQQSOs,
\item the RLQSOs are in  denser environments than RQQSOs. 
\end{itemize}

At low redshift, evidence points towards RLQSOs existing in denser
environments than RQQSOs \citep[e.g.][]{1990ApJ...348...38S}, although
these low-$z$ RLQSOs have  lower luminosities than those in our
sample. When compared to galaxies of the same mass, RLQSOs are
preferentially found in galaxies at the center of groups and halos
\citep{2007MNRAS.379..894B}. \citet{2007arXiv0709.2911K} find that the
local density of galaxies around RL-AGN is a factor of two higher than
around RQ-AGN matched in velocity dispersion, redshift and stellar
mass. All these results, combined with our excess of low-velocity
absorbers in the RL QSOs, might  suggest that RL QSOs exist in
denser environments. 

On the other hand, radio jets have been found to drive low-velocity
($\sim$100\,km/s) outflows of hydrogen \citep{2007arXiv0710.1189M}.  A
similar conclusion has been reached for high-redshift radio galaxies
by \citet{2007A&A...475..145N}. Hydrodynamical simulations show how
cool gas which may lead to absorption line systems can build up behind
the bow shock of radio jets \citep[e.g.][]{2002A&A...386L...1K}.  Such
a scenario should be testable, as it would imply a clear relation
between the presence of absorbers and the orientation of the jet.

\subsection{Absorber and QSO halo masses}

Our derived values of the cross-correlation lengths of QSOs with \civ\
and \mgii\ systems ($r_0 = 5.8\pm1.1$ and $r_0 =
5.0\pm0.4h^{-1}$\,Mpc, respectively) are similar to the correlation
lengths of the reddest and most massive galaxies at similar redshifts
($z\sim2$ and $z\sim1$ respectively).


\citet{2006A&A...452..387M} find $r_0$ values of 2.5 $h^{-1}$ Mpc for
blue galaxies and 4.8$h^{-1}$\,Mpc for red galaxies at $z\sim0.8$ in
the Vimos-VLT Deep Survey (VVDS).  \citet{2007arXiv0708.0004C} find
slightly higher values in the DEEP2 survey, 3.9 and 5.2$h^{-1}$\,Mpc
for blue and red galaxies at $z\sim1$. \citet{2005ApJ...619..697A}
measure $r_0=4.5\pm0.6, 4.2\pm0.5h^{-1}$\,Mpc and $\gamma=1.6$ at $z=1.7$ and $2.2$ for
galaxies selected through the Lyman-break technique.

The auto-correlation for QSOs at similar redshifts is less well
known: $r_0=6.5\pm1.6h^{-1}$\,Mpc and $\gamma=1.58\pm0.2$
\citep{2007AJ....133.2222S}; $5.55\pm0.29h^{-1}$\,Mpc and
$\gamma=1.63\pm0.054$ on large scales \citep[][ but see this paper for
caveats on the measured shape of the quasar auto-correlation
function]{2005MNRAS.356..415C}.

Taking the geometric mean of the QSO and galaxy auto-correlation
amplitudes assuming $r_{0}^{gg}=4.5$ and $r_{0}^{qq}=6h^{-1}$\,Mpc gives
$r_{0}^{gq}=5.2h^{-1}$\,Mpc, which matches our absorber-QSO cross
correlation amplitudes\footnote{The small difference in $\gamma$ makes
a negligible difference to this result.}.
Converting $r_0$ into the
alternative measure of comoving clustering length $\Delta_8$, and following \citet{2002MNRAS.336..112M}
results in QSO dark matter halo masses of $>10^{12.5}$M$_\odot$
at $z\sim2$ and $>10^{13}$M$_\odot$ at $z\sim1$. \civ\ and \mgii\
absorption line systems are then typically located in galaxies with
dark matter halo masses of $>10^{12-12.5}$M$_\odot$.

\subsection{Caveats and Limitations}

The primary limitation in the present analysis is the unknown
density distribution of absorber clouds in the immediate vicinity of
the QSO. Our toy model assumes a simple cut--off radius inside of
which no absorbers can exist, and beyond which power-law clustering
dominates. In a more realistic model, there might be variable cut--off radius
($R_{\rm cut}$) dependent on the density of the absorbing cloud. 
Such models are beyond the scope of the present analysis.

We note that a small offset remains between the line-of-sight
distributions predicted by our model and the the observed one.  The
magnitude of the observed offset is around 200\,km/s or 2.85\,Mpc, and
may be accounted for by the fact that galaxies in the vicinity of the
QSO will have a net motion caused by infall into potential well of the
QSO host halo.  This will alter the observed shape and peak position
of the absorber distribution. A full treatment of infall requires
detailed cosmological simulations, and will be addressed in future
work.

On the observational side, it is clear that a careful correction of
the QSO redshifts for the effect of the blue-shifting of emission lines
is warranted in order to constrain the detailed shape of the \civ\
absorber distribution. Follow-up observations in the near infra-red of
the \oii\ and \oiii\ emission lines in a relatively large sample of
high-$z$ QSOs would provide a very useful basis for measuring these
effects. The increase in number of QSOs from SDSS DR3 to SDSS DR7 will
also further enable us to constrain the precise shape of the velocity
distribution, and the fraction of dense absorbers which survive the
ionising radiation of the QSO.

\section{Summary}\label{sec:summary}
We have used a cross-correlation analysis of QSO-absorber pairs to
measure the strength of narrow absorber clustering around QSOs. A
simple model to convert the 3-D distribution of QSO-absorber
separations into a line-of-sight distribution in velocity space is
presented.  Our modelling allows us to reach the following conclusions
for \civ\ systems:

\begin{itemize}
\item Galaxies in the vicinity of the QSO may contribute as much as
  $\sim$55\% of the central spike of absorbers with
  $\Delta v\lsim0.01c$. The remaining $\sim$45\% can be directly linked to 
  the presence of the QSO.
\item A high-velocity tail of narrow absorbers is observed, which
  cannot be explained by galaxy clustering. These intrinsic absorbers
  make up $\sim$45\% of narrow absorption line systems with
  $0.01<\Delta v/c<0.04$.
\item The high-velocity tail is visible in both RQ and
  RL QSOs. RL QSOs show an enhanced number of absorbers at
  low velocities.
\end{itemize}

For \mgii\ absorbers we find:
\begin{itemize}
\item The spike of ``associated'' absorbers close to the QSOs in
  velocity space is consistent with being caused by galaxy clustering,
  but our analysis cannot rule out the possibility that a considerable
  fraction are caused by absorbers intrinsic to the QSO and its host.
\item The QSO destroys \mgii\ clouds well beyond the expected scale
  of its own halo, out to at least 800\,kpc (comoving). 
\item A high-velocity tail of intrinsic \mgii\ absorbers is detected,
  at a level of 20$\pm$5\% that observed in the \civ\ absorbers.
\end{itemize}

In the future, the larger absorber samples provided by later releases
of the SDSS survey data, improved methods for obtaining reliable QSO
redshifts, and investigations of ionisation and line width trends with
velocity, will contribute substantially to isolating the physical
processes responsible for QSO outflows detected through narrow
absorption lines.

\section*{acknowledgments}
We would like to thank Craig Hogan, Stuart Sim, Philip Best, Jeremy
Blaizot, Cheng Li, and Robert Brunner for useful discussions and
comments. This paper made use of the IDL MPFIT package by Craig
Markwardt http://cow.physics.wisc.edu/~craigm/idl/.

Funding for the SDSS and SDSS-II has been provided by the Alfred
P. Sloan Foundation, the Participating Institutions, the National
Science Foundation, the U.S. Department of Energy, the National
Aeronautics and Space Administration, the Japanese Monbukagakusho, the
Max Planck Society, and the Higher Education Funding Council for
England. The SDSS Web Site is http://www.sdss.org/.

The SDSS is managed by the Astrophysical Research Consortium for the
Participating Institutions. The Participating Institutions are the
American Museum of Natural History, Astrophysical Institute Potsdam,
University of Basel, University of Cambridge, Case Western Reserve
University, University of Chicago, Drexel University, Fermilab, the
Institute for Advanced Study, the Japan Participation Group, Johns
Hopkins University, the Joint Institute for Nuclear Astrophysics, the
Kavli Institute for Particle Astrophysics and Cosmology, the Korean
Scientist Group, the Chinese Academy of Sciences (LAMOST), Los Alamos
National Laboratory, the Max-Planck-Institute for Astronomy (MPIA),
the Max-Planck-Institute for Astrophysics (MPA), New Mexico State
University, Ohio State University, University of Pittsburgh,
University of Portsmouth, Princeton University, the United States
Naval Observatory, and the University of Washington.

\bibliographystyle{mn2e}



\end{document}